\newcommand*{\aaCit}{Astron. Astrophys.} % Astronomy and Astrophysics
\newcommand*{\anapCit}{Ann. d'Astr.} % Annales d'Astrophysique
\newcommand*{\aphCit}{Astropart. Phys.} % Astroparticle Physics
\newcommand*{\apjCit}{Astrophys. J.} % The Astrophysical Journal
\newcommand*{\arXivCit}{arXiv} % SISSA Database
\newcommand*{\JCAPCit}{J. Cosmolog. Astropart. Phys.} % Journal of Cosmology and Astroparticle Physics
\newcommand*{\mnrasCit}{Mon. Not. R. Astron. Soc.} % Monthly Notices of the Royal Astronomical Society
\newcommand*{\PASJCit}{Publ. Astron. Soc. Japan} % Publications of the Astronomical Society of Japan
\newcommand*{\PLSSCit}{Planet. Space Sci.} % Planetary and Space Science
\newcommand*{\SSRvCit}{Space Sci. Rev.} % Space Science Reviews
\newcommand*{\PhysRevCit}{Phys. Rev.} % Physical Review
\newcommand*{\PRLCit}{Phys. Rev. Lett.} % Physical Review Letters
\newcommand*{\RPPhCit}{Rep. Prog. Phys.} % Reports on Progress in Physics
\newcommand*{\NatureCit}{Nature} % Nature
\newcommand*{\ScienceCit}{Science} % Science
\newcommand*{\nodata}{...}

\documentclass[twocolumn]{article}
\usepackage[utf8]{inputenc}
\usepackage{amssymb}
\usepackage{graphicx}
\usepackage[english]{babel} % The pack­age man­ages cul­tur­ally-de­ter­mined ty­po­graph­i­cal and other rules.
\usepackage[square,numbers]{natbib}
\begin{document}

\title{Relativistic Slowing Down Shocks as Sources of GRB Lags}
\author{Janusz Bednarz}
\date{Uniwersytet Warmińsko-Mazurski w Olsztynie,\\
      Wydział Matematyki i Informatyki,\\
      ul. Żołnierska 14, 10-561 Olsztyn, Poland}

\maketitle

\label{firstpage}

\begin{abstract}
{
\noindent We demonstrate how slowing down ultrarelativistic shocks create gamma ray burst (GRB)
lags. Reflection process produces positive lags and Cracow acceleration produces negative lags.
We describe the process of Cracow acceleration and present two ways in which the seed
particles are injected into the upstream plasma. Strong decelerating shocks in the presence
of the acceleration processes account for the observed hard energy spectra of accelerated
electrons with spectral indices smaller than the value 2.2. We claim that during the strong
deceleration stage the rise-fraction of seed particles is formed upstream of the shock.
The rise-fraction feeds the Cracow surge and normal seed particles feed the reflection surge.
We present the model of the microphysics of relativistic plasma that aims at explaining
the required disturbances of the movement of particles upstream of the shock that allow
for Cracow acceleration. We show that Cracow acceleration can produce
ultra-high energy cosmic rays (UHECRs).}
\end{abstract}

\noindent{\bfseries Keywords:}

\noindent
{Acceleration of Particles; Cosmic Rays; Radiation Mechanisms: Non-Thermal; Shock Waves }

\section{Introduction}\label{sec:intro}
\citet{Fermi49} proposed a mechanism being supposed to explain cosmic rays acceleration.
The idea is analogous to the acceleration of a tennis ball. Interstellar clouds are tennis rackets
and a particle is the tennis ball. There is a small difference here, tennis rackets rarely
decrease the energy of the ball. Cosmic rays are accelerated in this mechanism according
to the second order Fermi acceleration method. It should be called Fermi acceleration.

The concept that shock waves could accelerate particles appeared slowly. It was foreshadowed by
\citet{Hoyle60} who postulated that shocks could efficiently accelerate particles but without
specifying a mechanism. \citet{Parker58} and \citet{Hudson65,Hudson67} attempted to obtain
such mechanism based on pairs of converging shocks and, most notably, \citet{Schatzman63}
constructed a theory based on perpendicular shocks where particles keep bumping into a hydromagnetic
shock front and increase their energy in the Fermi-like way.

The real acceleration mechanism was described in four seminal papers, \citet{Krymsky77},
\citet{Axford77}, \citet{Bell78a,Bell78b} and \citet{Blandford78}. Since many persons aspired to
the name of the discoverer, therefore this mechanism was called diffusive shock acceleration.
It is often called Fermi acceleration, but it is incorrect, since Fermi does not have anything
to do with particle acceleration at shocks. The name stuck to this mechanism
(see \citet{Hoshino08}, page 940) and we use it in this paper.

One should, by the way, recall nomenclatures that mislead uninformed readers about
a misunderstanding which persists in the literature. First order Fermi acceleration and second
order Fermi acceleration are not a real mechanism. It is the method which produces the Fermi
statistical dependence in which the proportional gain of energy is the same for all particles.
There are two or more media in the method that are moving with the average speed equal to $V$.
The speed of a particle is much larger that $V$ and is equal to $c$. The particle bounces
between the media and gains energy. If the direction of $V$ is always opposite to the direction of
$c$ in each collision then the average energy gain of the particle is proportional to $V/c$ and
that is first order Fermi acceleration. If the directions are random then the average energy gain is
proportional to $(V/c)^2$ and that is second order Fermi acceleration. The both methods appear
in separate phenomena and there is no need of applying a shared name for them. It could sometimes
lead to ambiguity. In rare cases when it is necessary one can use the phrase
'all Fermi acceleration processes' (\citet{Drury83}, page 987).
First order Fermi acceleration assumes that $V\ll c$. It implies that the distribution
of accelerating particles in the medium becomes nearly isotropic so that
the particles could reach the opposite medium. First order Fermi acceleration
predicts that the energy of accelerating particles increases $\sim \gamma^2$ times
in each cycle upstream-downstream-upstream if the mechanism is applied to relativistic shocks.

Some authors are trying to name the idea of the particle energy increase as a result of 
particles bouncing back and forth across the shock wave as Fermi acceleration, but it is
incorrect. It is Hoyle-Schatzman idea.

In Fermi acceleration particles can cross the shock repeatedly thanks to magnetic field
fluctuations downstream of the shock and subluminal shock geometry. Magnetic field fluctuations
upstream of the shock do not play significant role here, since the shock will always catch up with
the particle that is wandering there. This mechanism does not accelerate particles in superluminal
shocks (\citet{Bell78a}, \citet{Drury83}) and therefore is limited to non-relativistic
and poorly relativistic shocks, since relativistic shocks are generically superluminal.

For a lot of years an unresolved riddle existed in astronomy. Observations have shown that in areas
where relativistic shocks are present particles are accelerating effectively,
but nobody was able to find the mechanism.

The acceleration mechanism that operates at ultrarelativistic shock fronts was discovered
in Cracow (\citet{Bednarz98}). From now on we will call it Cracow acceleration since it
was discovered in Cracow. This important discovery was presented earlier at three conferences
\citet{Bednarz97a,Bednarz97b,Bednarz97c}.

\begin{figure}
\centering{\includegraphics[width=78mm]{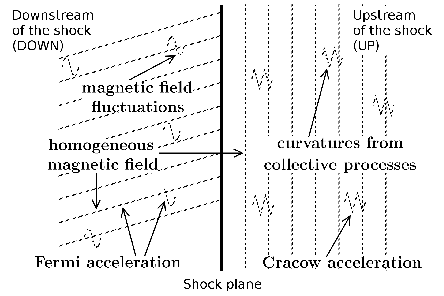}}
%\begin{center}%figure_1_1
%\FigureFile(78mm,){mechanism.eps}
%\end{center}%figure_01
\caption{Fermi acceleration arises due to inclination and fluctuations of the magnetic field
         downstream of the shock. Cracow acceleration arises due to the local magnetic and electric
         fields generated by collective processes upstream of the shock. }
\label{fig:mechanism}
\end{figure}

\citet{Bednarz98} showed that ultrarelativistic shocks accelerate particles and accelerate
them more effectively if their Lorentz factor grows. It results from the place and the way
the particles are getting the chance of return to the shock. In Fermi acceleration they return
to the shock because of conditions downstream of the shock and in Cracow acceleration because of
the disturbances of particle trajectories that come from the magnetic and electric fields
upstream of the shock that are generated by collective processes (Fig. \ref{fig:mechanism}).

There were a few proposals before, but their mechanisms are ineffective (because of steep spectra,
the slow acceleration time, the small maximum energy that can be achieved and the lack of seed
particles). \citet{Begelman90} proposed shock-drift acceleration at relativistic shocks
to operate at hot spots of extragalactic radio sources. Afterwards, \citet{Hoshino92}
described a process of shock acceleration of positrons to non-thermal distributions devoted
to account for the synchrotron radiation observed in the Crab Nebula and hot spots. Other proposals
are magnetic reconnection (e.g. \citet{Romanova92}), shock surfing acceleration
(e.g. \citet{Katsouleas83}) and MHD turbulence (e.g. \citet{Bieber94}).

Reflection process operates at relativistic shocks. It was known that a particle increases
its energy $\gamma^2$ times in the cycle upstream-downstream-upstream. \citet{Vietri95}
proposed repeated crossings at an ultrarelativistic shock to occur and argued that it would lead to
$\gamma^2$ energy gain per crossing cycle. \citet{Bednarz98} found that because of strong
anisotropy of cosmic rays upstream of the shock the gain equals to $\sim2$, what was confirmed
by \citet{Gallant99}. However, flowing in particles have isotropic distribution and gain $\gamma^2$
times energy in the first cycle and it was called reflection (\citet{Bednarz99}). In this paper
we show that reflection process is an important element of cosmic ray acceleration.

After \citet{Bednarz98} they have been trying to include the idea of cosmic ray anisotropy upstream
of the shock, but in the Fermi acceleration context, \citet{Kirk00}; \citet{Achterberg01};
\citet{Vietri03}, \citet{Lemoine03} - disordered shocks and \citet{Ellison02} - parallel shocks.
Fermi acceleration takes place at relativistic shocks in two cases not-appearing in the outer space,
at practically parallel shocks (subluminal geometry) and at shocks with completely disordered
magnetic field downstream of the shock (we named them disordered shocks). One would call these two
theoretical cases Peacock acceleration (\citet{Peacock81}). The Weibel instability is preventing
disordered and parallel shocks from being formed what observations of polarised emission
are confirming (\citet{Steele09}, \citet{Goetz09}).

\citet{Niemiec06a}, \citet{Niemiec06b}, \citet{Lemoine06} were trying to examine
the acceleration process and used 'realistic' magnetic field fluctuations upstream of the shock.
They have failed to get effective acceleration at superluminal shocks. \citet{Pelletier09}
discussed it analytically and have been found to agree with the numerical results.
Their lack of effective acceleration results from this that they have applied normal conditions
(types of fluctuation spectra) to the phenomenon which is completely unknown.

Cracow acceleration needs a type of the fluctuation spectrum where almost entire fluctuation
energy is gathered in very intense waves about a small length. The length could be so small as
the size of the atom. We do not believe that such fluctuations exist and propose the loop plasma
model as the solution to the problem in section \ref{sec:micro}.

We expect that astronomical sites where particles are accelerated in Cracow acceleration
are relativistic shocks with the Lorentz factor much larger than 1 and the helical (perpendicular
to the shock normal) perpendicular magnetic field produced by the Weibel instability
(\citet{Medvedev99}). In gamma ray bursts (GRBs) the process does occur not only
in prompt gamma-ray pulses, but also in X-ray flares (\citet{Margutti10}) and in precursors
(\citet{Burlon09}) where very similar spectral properties are observed. Long and short GRBs
are also showing the same acceleration process (\citet{Guiriec10}).

In pulsar wind nebulae (PWNe) where (contrary to GRBs) the plasma downstream of the shock is
approximately at rest with respect to ISM, not only leptons but probably protons are being
accelerated (\citet{Li10}). There is not only the same mechanism of the acceleration present, but
probably similar physical conditions, since the assumed particle injection spectrum in the form
of a broken-power law turns out to have spectral break at an similar energy for all sources
(\citet{Bucciantini11}).

We think that particles are accelerated in the mechanism of Cracow acceleration in gamma-ray
binaries (\citet{Cerutti09}), in X-ray binaries where relativistic jets are formed and in jets
thrown away by active galactic nuclei. An example of X-ray binaries is Cygnus X-3 where electrons
gain energy at the place where the jet is recollimated by the stellar wind pressure and forms
a shock (\citet{Dubus10}).

In the test particle limit the energy spectral index of accelerated particles is only dependent on
the compression ratio of the plasma through the shock by $\beta=(R+2)/(R-1)$
(\citet{Bell78a}, \citet{Drury83}). The formula is not valid for relativistic shocks but it is
good enough for the estimation how $\beta$ increases when the compression falls down.

The magnetisation parameter, $\sigma$, is equal to the upstream Poynting flux relative to the total
mass-energy flux (\citet{Appl88}, \citet{Kennel84a}). \citet{Kennel84a} presented the derivation of
the Rankine-Hugoniot relations for perpendicular shocks. They found that
$R=3(1-4\sigma)$ for $\sigma\lesssim 0.01$ and $R=1+1/(2\sigma)$ for $\sigma\gtrsim 10$.
We estimate that Cracow acceleration needs $\sigma<0.1-0.01$ to accelerate efficiently if one
applies normal plasma, but we devise the model of loop plasma (section \ref{sec:micro}) which could
have a larger value of $\sigma$ and the standard compression $R=3$.

\citet{Kirk03} have found that the Poynting flux can be dissipated before the pulsar wind
reaches the inner edge of the Crab Nebula. This is in accordance with the value of $\sigma=0.003$
near the termination shock of the nebula (\citet{Kennel84a}, \citet{Kennel84b}) which is far more
sufficient than Cracow acceleration requires.

The present paper is based on \citet{Bednarz12}. In the theory of cosmic rays acceleration
little has changed since then and observations do not point to one model.
Below, we update references in the field of research.

\citet{Kargaltsev15} provide a review of pulsar winds and PWNe, \citet{Buhler14}
a review of the Crab Nebula and its pulsar, \citet{Kagan15} a review of the physics
of magnetic reconnection and \citet{Granot15} a review of GRBs.
\citet{Lemoine15} showed, taking into account Fermi acceleration, that known pulsar
wind nebulae are not powerful enough to confine ultra-high energy cosmic rays (UHECRs).
\citet{Sironi15} in a review of the physics of relativistic shocks show
that particle acceleration, when efficient, modifies the turbulence around the shock
and wonder how it influences further acceleration.

The context of the paper is organised as follows. In section \ref{sec:micro} we present the model
of the microphysics of relativistic plasma as well as the way the simulations of the relativistic
plasma should be conducted. In section \ref{sec:Cracow} the mechanism of Cracow acceleration and
in section \ref{sec:simulations} the numerical simulations are described. In section
\ref{sec:slowing_down} we give thought to the influence of the deceleration of the shock on
the energy spectra of accelerated particles. Results of simulations are presented in section
\ref{sec:results}, there are many tables and figures. Section \ref{sec:Fermi} is devoted to
the problem of conditions that enable Fermi acceleration to produce negative lags. In section
\ref{sec:two_processes} we postulate the formation of the rise-fraction of particles upstream
of the shock at the strong deceleration stage. The rise-fraction feeds the Cracow surge and normal
seed particles feed the reflection surge. In section \ref{sec:UHECRs} we show that Cracow
acceleration can produce UHECRs. The seed particles problem is discussed in section
\ref{sec:seed_particles}. We summarise and discuss our paper in section \ref{sec:summary}.

\section{The microphysics of relativistic flows}\label{sec:micro}
Relativistic plasma is an obscure phenomenon. Magnetic waves are too weak to disturb the particle
movement so strong as Cracow acceleration requires. There must be some collective processes
that build magnetic and electric fields on a micro-scale.

One should remember that current numerical simulations and analytic models are useless in
the topic of relativistic plasma since they do not reach the required accuracy.
We will explain this opinion with the use of two examples of PIC simulations.

\citet{Spitkovsky08} simulated relativistic shocks in 2D and had used an artificial wall
that produced an artificial returning stream. Collective processes in plasma are sensitive
to global changes of plasma and the stream is preventing the processes from coming into existence.
The author has detected particle acceleration but has not given any reason for this. In our opinion
the artificial returning stream accelerates particles.

\citet{Nishikawa06} performed simulations with $\sim3.8\cdot10^8$ particles, but we have
counted $\sim2.5\cdot10^8$ what is unimportant here. The important thing is that $1.25\cdot10^8$
hydrogen atoms have mass $2.1\cdot10^{-16}g$, $1.25\cdot10^8$ electron-positron pairs have mass
$1.1\cdot10^{-19}g$, a flu virus has mass $7\cdot10^{-13}g$ and typical plasma flows of GRBs have
rest mass $\sim10^{27}g$ if we apply canonical energy release of $\sim10^{51}erg$, the Lorentz
factor of the plasma equal to $1000$ and neglect the Poynting flux.

The authors have scaled their simulations to density of the interstellar medium in order to get
the length of the simulation box equal to $4\cdot10^7cm$ and have not scaled to density of the tail
of a comet, for example, which is $10^5$ times higher.
The authors have compared the length of the simulation cell to the electron skin depth in order
to show that the shock fits in the simulation box, but it is untrue. The thickness of the shock
is comparable to the gyroradius of a thermal ion (\citet{Drury83}) and it compares unfavourably
with the size of their simulation box. The authors have not reached dynamic equilibrium so that
they do not know if their results are stable.
The authors have written, '.. the ‘flat’ (thick) jet fills the computational domain in
the transverse directions (infinite width). Thus, we are simulating a small section of
a relativistic shock infinite in the transverse direction.' and that is untrue again.
They have applied some boundary conditions and named them 'thick jet' or 'infinite width'.
All boundary conditions are artificial and can produce anything. Further below, we
propose how to manage with the problem.

We have shown that the current PIC simulations of relativistic plasma are pointless. They have
provided two intuitive hints only. First, relativistic plasma could produce seed particles
(\citet{Nishikawa03}, \citet{Hededal04}) and second, the plasma could form current filaments
(\citet{Medvedev99}). We believe that the task of creation of relativistic plasma in PIC
simulations from scratch is outside present capabilities of mankind. We come up with an idea
and show the way the PIC simulations should approach the problem of relativistic plasma.

\begin{figure}
\centering{\includegraphics[width=78mm]{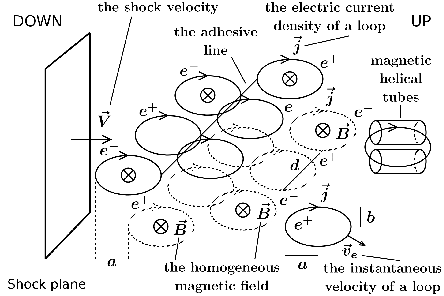}}
%\begin{center}%figure_2_1
%\FigureFile(78mm,){plasma_e-e+.eps}
%\end{center}%figure_02
\caption{The model of loop plasma. The six solid line ellipses represent the upper layer and
         the six dashed line ellipses the bottom layer. The detailed description is in text.}
\label{fig:ep_plasma}
\end{figure}

At first, we advance a hypothesis on the microphysics of relativistic plasma. The hypothesis aims
at explaining the required disturbances of the movement of particles upstream of the shock that
allow for Cracow acceleration. Now, we restrict our considerations to the electron-positron plasma.
Our model is presented in Fig. \ref{fig:ep_plasma}. The Weibel instability forms electric current
filaments in the shape of an oval which we substitute for an ellipse. Sometimes, we are calling
this oval the loop. The sense of the electric current density $\vec{j}$ is the same for all loops.
Each loop contains identical electric charges, electrons or positrons, which are indicated
in Fig. \ref{fig:ep_plasma} by $e^-$, $e^+$, respectively. The particles are arranged evenly
along the loop in this way that the centrifugal force and the electric force are balanced by
the Lorentz force. The isolated loop in the shape of a circle is in unstable equilibrium and
the environment flattens it into the oval. We have no idea which oval is more stable, the one
elongated parallel or perpendicular to the shock normal, but PIC simulations suggest that
ovals elongated parallel to the shock normal make relativistic plasma
because current filaments tend to be parallel to the direction of the flow.

The system presented in Fig. \ref{fig:ep_plasma} is a simple model only. A real system is stable
and the Weibel instability rebuilds continuously the ovals.
The system is certainly not so perfect as presented in Fig. \ref{fig:ep_plasma}. The ovals have
different diameters, they oscillate, have global movements (the instantaneous velocity of a loop is
$\vec{v}_e$). The ovals are being distorted in space. There is a fraction of particles that do not
belong to any oval (free particles).

The ovals form the homogeneous magnetic field - $\vec{B}$. They are divided into layers which are
parallel to the shock normal. Each layer is divided into rows which are perpendicular to the shock
normal. Ovals along a row are arranged alternately, the electron oval follows the positron oval and
vice versa. We adopt the rule that the ovals approach the geometry that minimise the magnetic field
opposite to $\vec{B}$. We have not carried out any calculations, but we think that two adjacent
rows adhere to the shared adhesive line. We indicate lengths of semi-axes of the ellipse by $a$ and
$b$, the distance between two consecutive ellipses along the row by $d$. Each layer is translated
in regard to the symmetrical position of the adjacent layer for the length of $a$ parallel
to the shock normal and $d/2$ perpendicular to the shock normal. The distance between two adjacent
layers depends on values of $a$, $b$ and $d$. The lengths of $a$, $b$, $d$ are comparable.

Loop plasma consists of particles in loops (we name them loop particles) and free particles. Loop
particles correspond to thermal particles of normal plasma. The part of free particles with
energies comparable to the energy of loop particles or lower correspond to thermal particles also.
The free particles that have energies a few times larger than loop particles form the fraction of
seed particles. Seed particles gain their energy by accelerating in the electric field of loops. 

Even though the Lorentz factor of the shock is so small as $\sim5$, the gyroradius of
the particles at the onset of Cracow acceleration is more than $\sim 100$ times larger than
the gyroradius of loop particles. As a consequence, the non-resonant interaction between loops
and accelerating particles is kept.

In our model not only the magnetic helical tubes (Fig. \ref{fig:ep_plasma}) but also separation
of electric charges disturb cosmic rays. It is easy to prove whether such disturbances enable
cosmic rays to accelerate in Cracow acceleration. One should track trajectories of high-energy
particles upstream of the shock (Fig. \ref{fig:ep_plasma}) with their initial momenta nearly
parallel to the shock normal. The change of the direction of the momenta must be small while
the tracking. The momenta must be all the closer to the shock normal the Lorentz factor
of the shock is larger. Now, one derives the value of $Q_0$ from the tracking and puts it into
(\ref{eq:with_g}). If the derived energy spectral index is close to the asymptotic
energy spectral index then the system accelerates particles in Cracow acceleration.
Equation (\ref{eq:with_g}) is rough because our approach does not include loop plasma.
However, it is good enough to estimate whether Cracow acceleration works.

The proton-electron plasma is more difficult to analyse. We think that it consists of proton
loops and electron loops. Diameters of the electron ovals are much smaller then proton ovals.
The ovals also approach the geometry that minimise the magnetic field opposite to the homogeneous
magnetic field. Electron ovals tend to be close to the electric current of proton loops.

We name the described plasma loop plasma. The distribution of particle velocities in loop plasma
is not the Maxwellian one therefore the temperature is not defined in the usual meaning.
It is hard to say what the downstream plasma will look like if loop plasma comes through the shock.
However, it is not important for Cracow acceleration which will work anyway.
The gradients of the velocity and the magnetic field of downstream plasma of slowing down shocks
influence reflection process slightly because the reflected particles have small gyroradius and
influence Cracow acceleration slightly because the change of the particle direction that allows
for the return to the shock takes place upstream of the shock. The influence grows with growing
deceleration.

The PIC simulations are useless unless one is able to detect dynamic equilibrium of plasma.
Creating relativistic plasma from scratch is outside the available computing power at the moment.
The PIC simulations must hunt for dynamic equilibrium of plasma rather than try and create
it from scratch. Below, we describe how the simulations should be conducted. We will demonstrate
how to catch dynamic equilibrium of electron-positron loop plasma. At first, we take into
account the upstream loop plasma only. The shock is far enough and does not affect the system.

One must use many processors at the same time (the parallel programming). The plasma is divided
into boxes. The boxes are within the space of a cuboid (the central cuboid). Each electric current
loop is within one loop box. Inside the loop box there are free particles also. Loop boxes must not
overlap so that their linear dimensions are changing. If necessary, a few loops can be inside one
box. The loop boxes can merge, split, disappear and be formed. Additionally, the plasma is divided
into free boxes that contain the remaining free particles inside empty spaces. All free boxes are
the same and fill up the cuboid entirely, they do not change.

To save the computing power, one processor handles many loop boxes or free boxes during one
computational step, but not at the same moment. The number of all boxes should exceed the number
of processors ten times for example. The electromagnetic force inside a box is calculated exactly,
the force from nearby boxes is also calculated exactly, but the force from any other box is taken
from a pattern database.

We replace the boundary conditions with a pattern database of the electromagnetic field of
pattern cuboids. The pattern cuboids surround the central cuboid up to a distance. There are
no empty spaces inside the system. The distance must be large enough to not change results
of simulations if it increases. If necessary, global electric and magnetic fields that
represent the electromagnetic field of plasma outside the distance is added. Sizes of
pattern cuboids can be different and the results can not depend on the sizes.

The biggest challenge is creating the pattern database (that includes boxes and cuboids) and
the efficient use of it. Each pattern in the database must have many items of accuracy. Let us say,
we have 15 items of accuracy and number 0 represents the lowest accuracy. Results of simulations
must be the same if we apply items numbered from a certain number up to number 14.

The pattern database changes as simulations bring closer to the dynamic equilibrium target.
The size of the central cuboid must be large enough to detect statistically independent pieces
of it. Pattern cuboids will be formed by splitting and merging the central cuboid in order to get
different sizes of them.

Plasma inside the central cuboid moves with a velocity in the observer's frame. The plasma is
supplied with boxes at one end of the central cuboid. The boxes disappear at the opposite end.
The entering boxes are changing. They are replaced with boxes similar to those ones nearby
the opposite end, but not too close to the boundary of the central cuboid. The simulations must run
until all boxes inside the central cuboid are statistically homogeneous, pattern cuboids are
similar to the central cuboid and the conditions do not change.
It is the dynamic equilibrium of plasma.

The next stage is to catch equilibrium of the plasma around the shock. It is easy because
we know what the entering boxes look like. The central cuboid must contain the upstream plasma,
the shock and the downstream plasma. Initial boxes of the shock and downstream plasma must resemble
the shock conditions. Simulations are conducted in the shock rest frame.

The equation of motion of the shock must be known from astronomical observations in order
to describe loop plasma of the slowing down shock. Simulations should be conducted in the shock
rest frame. The entering boxes are supplied according to the equation of motion. The simulations
will start with the initial Lorentz factor of the shock and loop plasma of the constant velocity
shock of the same Lorentz factor. Initial loop plasma will be replaced with loop plasma produced
at a lower Lorentz factor and the factor will be increased until the factor close to the initial
Lorentz factor will be reached.

We have presented the sketch of simulations of relativistic plasma. We do not know the specific
structure of loop plasma and the required number of boxes and cuboids that allows to capture
the collective processes but we think that one needs $10^5-10^7$ processors to perform the task.
It requires much of intellectual efforts also, but it has to be done if we wish to figure out
the microphysics of relativistic plasma.

The issue still remains to be explained whether the plasma that allow for Cracow acceleration
exists. The answer is simple, observations resolve the problem. Cracow acceleration predicts a lot
of peculiar facts as superluminal shocks acceleration, acceleration at shocks with extremely large
Lorentz factors (PWNe), the negative GRB lag, the extremely short acceleration time and others.

Determining the acceleration time in PWNe would be the crucial proof. The plasma downstream of
the shock in pulsars is at rest with respect to the interstellar medium (ISM) and the value of
the homogeneous magnetic field can be detected there. The acceleration time in the mechanism of
Cracow acceleration is equal to 1 or is shorter if measured in units of the homogeneous magnetic
field downstream of the shock (\citet{Bednarz00}). Mechanisms in which the change of the direction
of the movement of a particle enabling it the return to the shock takes place downstream
of the shock can not achieve such short acceleration time.

\section{The mechanism of Cracow acceleration}\label{sec:Cracow}
In this paragraph we follow simulations of \citet{Bednarz00} and \citet{Bednarz04}. We will
sometimes denote '(the plasma) upstream of the shock' as UP and '(the plasma) downstream
of the shock' as DOWN. We have simulated relativistic shocks with Lorentz factors $\gamma\gg 1$ and
the perpendicular homogeneous magnetic field. Downstream of the shock the pure homogeneous magnetic
field is present, and upstream of the shock there are additional disturbances of the particle
movement caused by collective processes. The disturbances pattern UP is characterised by a value of
the parameter $Q$ that fulfils the formula from \citet{Bednarz04}

\[\frac{\kappa_\perp}{\kappa_\parallel}=\frac{1}{1+Qp^2}\, ,\]
where $p$ is the particle momentum and $\kappa_\perp$, $\kappa_\parallel$ are perpendicular
and parallel diffusion coefficients, respectively.

The formula is based on the formula from \citet{Jokipii70},
$\frac{\kappa_\perp}{\kappa_\parallel}=\frac{1}{1+(\omega_c\tau)^2}$,
where $\tau$ is the typical scattering time and $\omega_c$ is the cyclotron frequency.
\citet{Jokipii70} got it from \citet{Axford65} and \citet{Quenby66}.

We assumed that each scattering (it can be the sum of many uncorrelated scatterings) shifts
the direction of the particle momentum, $p$, at a small angle of $\sim 1/p$.
The assumption is good enough for loop plasma because the interaction is non-resonant and
the scatterings originate with the electric and magnetic fields of nearby loops mainly.
The force responsible for them is the Lorentz force,
$\vec{F}=e_0\vec{E}+e_0(\vec{v}\times\vec{B})$. The particle charge, $e_0$, and the velocity,
$|\vec{v}|\simeq c$, are always the same. We think that loops should be stable as a whole
during the interaction. The relativistic mass of the accelerating particle is much larger
than the mass of one loop particle (excluding a proton loop interacting with leptons).
Therefore, a loop particle can see the accelerating particle always the same, independently
of the momentum of the accelerating particle. We think that the loop changes as a whole
during the interaction and the changes of its electric ($\vec{E}$) and magnetic ($\vec{B}$)
fields do not depend on the momentum of the accelerating particle significantly. At least,
the difference should be small enough to allow the approximation $\sim 1/p$ to be correct.

We think that the uniform particle momentum scattering within a cone with a small angular
opening (\citet{Ostrowski91}) and the approximation $\sim 1/p$ (\citet{Bednarz04}) is a good
choice if we do not know the exact form of loop plasma. However, the real interaction
between accelerating particles and loop plasma is much more complicated. The loops are
elongated and therefore one should use the complete diffusion tensor rather than
$\kappa_\perp/\kappa_\parallel$. Most probably, some directions are favoured and momentum
changes are not evenly distributed on the small angular opening.

The numerical calculations involve particles with momenta systematically increasing over several
orders of magnitude. In order to avoid any energy dependent systematic effect we consider
the situation with all spatial and time scales – defined by the diffusion coefficient,
the mean time between scatterings and the shock velocity – proportional to the particle gyroradius
$r_g$, i.e. to its momentum. It means that the results are momentum independent and can be easily
scaled to any momentum. In particular we will be using the parameter $Q_0=Qp^2$ that determines
the disturbances pattern UP independently of the value of the momentum.

We have applied the light velocity, $c$, equal to 1 and the time is measured in $r_g/c$ units.
The plasma UP is hot and the compression is equal to $3U_s^2$, where $U_s$ is the shock velocity
in the upstream plasma rest frame.
We have chosen hot plasma because of subshocks in GBSs as the leading example but
a choice of the cold electron-proton plasma, the hot electron-positron plasma etc.
is unimportant because the compression differs only slightly. Moreover, we postulate
loop plasma where the compression is unknown at present.
The Lorentz factor of the plasma DOWN as seen UP, $\gamma_{ud}$,
can be approximated through the formula $\gamma_{ud}=0.71\gamma$. 

The mean time the particle spends UP between shock crossings, $\Delta t_u$, is measured in
the plasma rest frame. For better understanding we give $\Delta t_u$ in radians, for such an angle
the particle will turn in the wandering in the homogeneous field. The value is the same as
in $r_g/c$ units.

The mean change of the particle momentum direction between the moment when the particle is entering UP
and the moment when it is leaving UP is equal to $\Delta\alpha_u$ if measured in the plasma rest frame
and $\Delta\alpha_d$ if measured in the downstream plasma rest frame. Since upstream of the shock
particles momenta are directed almost parallel to the shock velocity we assume that
$\Delta\alpha_d\simeq 2\gamma_{ud}\Delta\alpha_u=1.42\gamma\Delta\alpha_u$.

Let us consider the particle that crosses the shock from UP to DOWN (UP-DOWN) with parameters $\phi$
and $r$, where the particle phase $\phi\in(0, 2\pi)$ is the angle between the direction DOWN-UP and
the projection of the particle gyroradius onto a plane that is perpendicular to the homogeneous
magnetic field, $r$, see \citet{Bednarz99}. The sense of the angle $\phi$ is the same as the sense of
rotation of the particle, and $r, \phi$ are measured in the downstream plasma rest frame.

The particle will cross the shock again if there is a solution of the equation

\[r\sin(\frac{t}{2}+\phi)\sin\frac{t}{2}+\frac{t}{3}=0, \]
at positive time $t$.

\begin{figure}
\centering{\includegraphics[width=78mm]{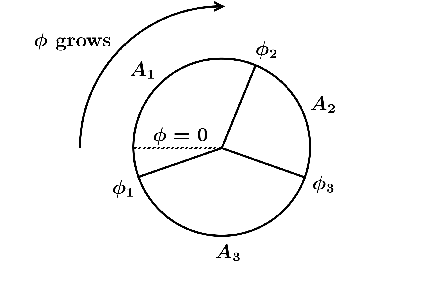}}
%\begin{center}%figure_3_1
%\FigureFile(78mm,){sets_of_parameters.eps}
%\end{center}%figure_03
\caption{The angle $\phi$ is increasing during the movement of the particle. The particle is
         entering UP with parameters from the set $A_3$, and is leaving UP with parameters
         from the set $A_1$ or $A_2$.}
\label{fig:sets_of_parameters}
\end{figure}

We will therefore distinguish three ranges of particle parameters (see Fig.
\ref{fig:sets_of_parameters}), $A_1=\{\phi : \phi\in(\phi_1, \phi_2)\}$ - the equation has not
the positive solution and the particle does not return to the shock,
$A_2=\{\phi : \phi\in(\phi_2, \phi_3)\}$ - the equation has the positive solution and the particle
returns to the shock, $A_3=\{\phi : \phi\in(\phi_3, \phi_1)\}$ - the particle with parameters from
this range crosses the shock DOWN-UP and can not cross the shock in the opposite direction,
$r$ sets the amount and determines $\phi$ values.

Values of $\phi_1$ and $\phi_3$ are given by analytic formulae,
$\phi_1=2\pi-\arcsin(1/(3r)),\,\,\, \phi_3=3\pi/2-\arccos(1/(3r))$, and $\phi_2$ must be calculated
numerically. A particle with $r<1/3$ does not return to the shock. Chosen values of parameters are
$r=1, \phi_1=5.94, \phi_2=1.96, \phi_3=3.48$; $r=0.5, \phi_1=5.55, \phi_2=2.96, \phi_3=3.87$;
$r=1/3, \phi_1=\phi_2=\phi_3=\frac{3\pi}{2}$.
  
The change of the particle direction, $\Delta\alpha_u$, consists of two parts, the change in
the process of wandering in the homogeneous magnetic field, $\Delta\omega_h$, and the change as
the result of collisions with disturbances, $\Delta\omega_f$, so we can write
$\Delta\alpha_u=\Delta\omega_h+\Delta\omega_f$.
We have neglected addition of vectors in this section.

A particle enters UP having parameters $(r, \phi)$ from the set $A_3$. If there are
no disturbances UP, the particle returns DOWN having $(r, \phi)$ from the set $A_1$. That is
the case of Fermi acceleration, where the acceleration is forbidden in superluminal shocks and
$\Delta\alpha_u=\Delta\omega_h$.

In Cracow acceleration $\Delta\omega_f$ plays the key role. Disturbances must be heavy so that
$\Delta\omega_f$ is large enough. The angle change $\Delta\omega_h$ is directing a particle
this way so that it gets parameters from the set $A_1$ and $\Delta\omega_f$ is directing randomly,
enabling the particle very much often changing $\phi$ from $A_3$ to $A_2$ through $\phi_3$.
If $\Delta\omega_f$ dominates then half of the particles change $\phi$ to a value from
the set $A_2$ and half from the set $A_1$. It implicates the short acceleration time equal to $1$.

The change of the particle direction through $\Delta\omega_h$ is the linear process because
$\Delta t_u$ is small for large $\gamma$, and through $\Delta\omega_f$ is the diffusion process,
and that is expressed by the formulae $(\Delta\omega_f)^2=C_f\Delta t_u$ and
$\Delta\omega_h=\Delta t_u$, where $C_f$ is constant. It yields the relation
$\Delta\alpha_d=1.42\gamma(\sqrt{C_f\Delta t_u}+\Delta t_u)$. If $\gamma$ grows then $\Delta t_u$
decreases. One can see now that the relative $\Delta\omega_f$ contribution to $\Delta\alpha_d$
grows with $\gamma$, which means that the particle energy spectral index,
$\beta$, decreases. The index would increase but only if $\Delta t_u$ decreased more quickly than
$1/\gamma^2$, but it does not occur as is shown below.

Let us return to the simulations. We have performed simulations for 14 $\gamma$ values,
$\gamma=2^n$, where $n=2,3,..,15$ and 500 $Q_0$ values from $5\cdot10^{-1}$ to $5\cdot10^7$
in logarithmic space. For larger $\gamma$ we started the computations from larger $Q_0$ than
$5\cdot10^{-1}$, but being small enough in order to get the satisfying range of the asymptotic
energy spectral index $\beta_0=2.23$ (\citet{Bednarz04}).

\begin{figure}
\centering{\includegraphics[width=78mm]{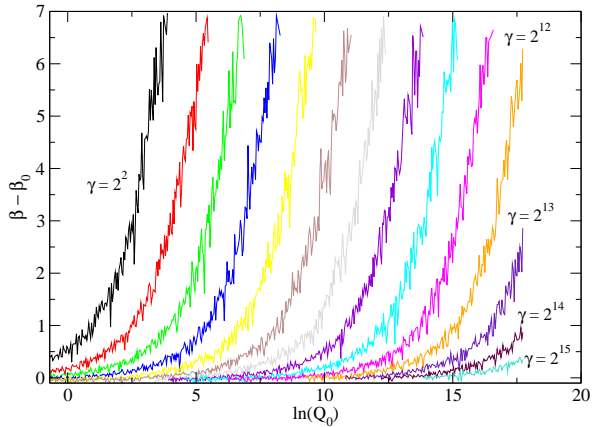}}
%\begin{center}%figure_3_2
%\FigureFile(78mm,){qs_raw.eps}
%\end{center}%figure_04
\caption{The energy spectral index of accelerated particles versus $Q_0$ for shock Lorentz factors
         in the range 4-32768.}
\label{fig:qs_raw}
\end{figure}

\begin{figure}
\centering{\includegraphics[width=78mm]{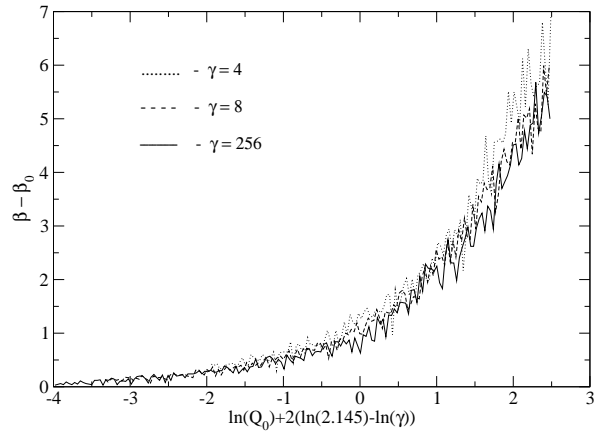}}
%\begin{center}%figure_3_3
%\FigureFile(78mm,){qs_fit.eps}
%\end{center}%figure_05
\caption{The same curves as in Fig. \ref{fig:qs_raw}, but converted and cut for the purpose of
         fitting. The curve for $\gamma=4$ (dotted line) is standing out a bit from the fitting and
         for $\gamma=8$ (dashed line) very little. Other curves fulfil the fitting closely and
         are represented by the curve for $\gamma=256$ (solid line).}
\label{fig:qs_fit}
\end{figure}

The obtained $\beta-\beta_0$ values as a function of $\ln(Q_0)$ are shown in Fig.
\ref{fig:qs_raw}. Individual curves represent constant $\gamma$ values and one can see that they
are similar. We have converted the curves according to the formula
$W=\ln(Q_0)+2(\ln(2.145)-\ln(\gamma))$. After the transformation the curves have become identical,
but the one for $\gamma=4$, and the curve for $\gamma=8$ has been standing out from this relation
very slightly (see Fig. \ref{fig:qs_fit}). The relation $\gamma^2/Q_0=const$ for constant $\beta$
is exact and there must be a simple theoretical explanation for it.


In order to get errors of fitting we have limited the range of the data to $W\in(-4, 2.5)$,
see Fig. \ref{fig:qs_fit}. We have excluded points for which $\beta\leq\beta_0$ and curves
for which $\gamma=2^{14}, 2^{15}$ because their $\beta$ values do not cover the upper range of $W$.
We have been fitting the relation $\beta-\beta_0=A\exp(BW)$ and have gotten $A=1.007\pm0.004$,
$B=0.722\pm0.003$. Finally, we have received the important relation

\begin{equation}
(\beta-\beta_0)^{1.39}=Q_0\left(\frac{2.145}{\gamma}\right)^2,\,\, \gamma>8.\label{eq:with_g}
\end{equation}

The numerical precision prevented us from carrying out calculations for $\gamma$ larger than
32768, but we can not see an obstacle that the above formula is not valid for all $\gamma>8$.

We have repeated the above procedure for accurate values of $\gamma_{ud}$ instead of $\gamma$
and have gotten the same results but the values of constants. Two first curves
($\gamma\in\{4,8\}$) the same diverge from the relation as for $\gamma$. The separation
of the curve for $\gamma=4$ depicts where Cracow acceleration is starting weakening,
for $\gamma=2$ it turns off practically.

\begin{figure}
\centering{\includegraphics[width=78mm]{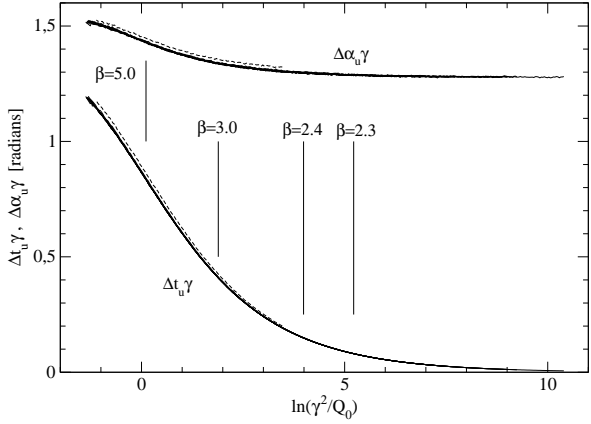}}
%\begin{center}%figure_3_4
%\FigureFile(78mm,){time_angle.eps}
%\end{center}%figure_06
\caption{The values of $\Delta t_u\gamma$ and $\Delta\alpha_u\gamma$ vs. $\ln(\gamma^2/Q_0)$.
 There are two sets of 14 curves each for constant $\gamma=2^n$, where $n=2,3,..,15$,
 in the figure. Curves for $\gamma=4$ stand out a bit and are represented by dashed lines.
 Other curves are merging into one curve for $\Delta\alpha_u$ and one for $\Delta t_u$.
 One can see from the figure that for large $\gamma$ the value of $\Delta\alpha_u$ decreases
 linearly with $\gamma$ and $\Delta t_u$ faster than that.}
\label{fig:time_angle}
\end{figure}

\begin{figure}
\centering{\includegraphics[width=78mm]{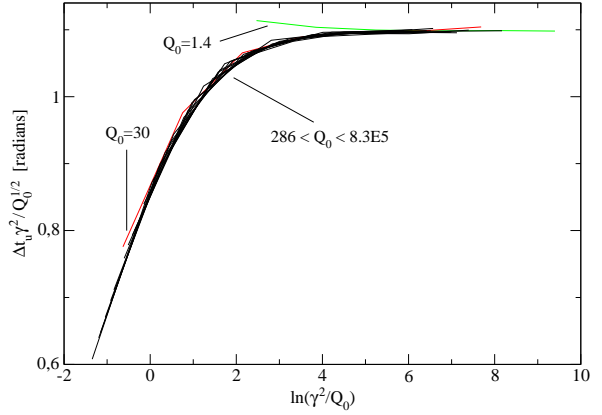}}
%\begin{center}%figure_3_5
%\FigureFile(78mm,){tgam2.eps}
%\end{center}%figure_07
\caption{The value of $\Delta t_u\gamma^2/\sqrt{Q_0}$ vs. $\ln(\gamma^2/Q_0)$.
 There are 14 curves for constant $Q_0$ in the figure. Curves for $Q_0=1.4$ and $Q_0=30$
 are indicated and other curves are merging. It is seen that $\Delta t_u\gamma^2/\sqrt{Q_0}$
 is constant for large $\gamma$.}
\label{fig:tgam2}
\end{figure}

Further, we have checked how $\Delta t_u$ and $\Delta\alpha_u$ are changing with $\gamma$.
The choice of $\ln(\gamma^2/Q_0)$ as independent variable enabled us to put curves in one line.
In Fig. \ref{fig:time_angle} 14 curves for constant $\gamma=2^n, n=2,..,15$ and in Fig.
\ref{fig:tgam2} 14 curves for constant $Q_0\in\{1.4, 30, 297, .., 8.3\cdot10^5\}$ are presented.
The curves for $\gamma=4$ (dashed lines) and $Q_0=1.4$ are diverging from general relations.


From Fig. \ref{fig:time_angle} it is seen that $\Delta\alpha_u\gamma$ is constant for large
$\gamma$ and $\Delta t_u$ decreases with $\gamma$ faster than $1/\gamma$. Moreover, we have put
lines that are indicating values of $\beta$ for fixed $\gamma^2/Q_0$ and now one can estimate
how reducing the contribution of $\Delta\omega_h$ to $\Delta\alpha_u$ is decreasing $\beta$
to $\beta_0$.

Fig. \ref{fig:tgam2} shows precisely how $\Delta t_u$ is changing with $\gamma$.
It decreases with $\gamma$ always slower than $1/\gamma^2$ and approaches the asymptotic value of
$1/\gamma^2$ for large $\gamma$. It guarantees that Cracow acceleration does not fade with growing
$\gamma$. For large $\gamma$, we have fitted the relation

\begin{equation}
\Delta t_u\gamma^2/\sqrt{Q_0}=1.097\pm0.002.\label{eq:dtu}
\end{equation}

Cracow acceleration could weaken with growing $\gamma$ because of the increasing energy of
reflected particles, but it is not. Let $p_s$ be the momentum of a seed particle. Reflection
process increases particle energy $\sim\gamma^2$ times, so that $Q_0\sim Q(p_s\gamma)^2$ at
the onset of Cracow acceleration process, so $1/(Qp_s^2)\sim\gamma^2/Q_0=const$ if $\beta=const$.
It means that the initial value of $\beta$ depends on plasma conditions upstream of the shock only
($Q$ and $p_s$). In real physical conditions this important result, similarly as previous results,
can be altered by magnetic field fluctuations downstream of the shock.

Cracow acceleration could weaken with growing $\gamma$ as a result of reducing the effectiveness of
reflection process, but it is a quite different problem for which plasma conditions DOWN should be
examined.

At the end of this section, we address the problem of the number of loops the accelerating
particle is interacting with. The seed particle momentum, $p_s$, is $k$ times larger than
the loop particle momentum. At the onset of Cracow acceleration the particle gyroradius
is $k\gamma^2$ times larger than the gyroradius of a loop particle and the particle
interacts with $m\simeq \Delta t_u k\gamma^2/2$ loops. We apply $\Delta t_u$ from (\ref{eq:dtu})
and get $m\simeq k \gamma^2 p_s\sqrt{Q}/2$. It determines that the accelerating
particle could interact with a part of a loop or possibly better a pair of loops because
the particle moves more often between two loops.

The first question is, does the particle accelerate in this case? Separate calculations should
be performed to answer this question correctly but the answer is yes if $m>0.5$. The conclusion
is drawn from the fact that if many loops are able to change the particle direction to allow for
Cracow acceleration then each loop does the same but better. It results from previous
discussions about the diffusion process and the linear process. If $m<0.5$ then
$\gamma^2 p_s\sqrt{Q}\lesssim 1/k$. This is the case of the extremely disordered magnetic field
which accelerates particles. We expect that loop plasma looks like this field if $m<0.5$.

The second question is, is it still Cracow acceleration? The answer is yes.
Cracow acceleration is the process that changes the particle direction UP in such a way that
the particle returns to the shock frequently even if DOWN is the pure perpendicular magnetic
field only. There is no problem with the beginning of Cracow acceleration.

\section{Simulations}\label{sec:simulations}
Below, the light velocity is used as the velocity unit, $c = 1$. As the considered particles are
ultrarelativistic ones, $p = E$, we often put the particle momentum for its energy.

In the simulations we consider the mean magnetic field configuration perpendicular to the shock
normal and disturbances of the particle movement caused by collective processes or magnetic field
fluctuations (weak disturbances only). We restrict our consideration to the test-particle
approximation in which it is assumed that particles are scattered by scattering centres in
the fluid but have no effect either on the fluid velocity or on the density of scattering centres.
Between two successive scatterings the particle is assumed to proceed along the undisturbed path
in the mean field. We model particle trajectory perturbations by introducing small-angle
random momentum scattering along the mean field trajectory
(\citet{Ostrowski91}, \citet{Bednarz96, Bednarz98}, \citet{Bednarz04}).

The perturbed magnetic field represents the traditional picture based on the concept
of magnetic scattering centres. It is simulated by the small amplitude particle momentum
scattering within a cone with angular opening $\Delta\Omega$ less than the particle
anisotropy $\sim 1/\gamma$, where $\gamma$ is the Lorentz factor of the shock. The particle
momentum scattering distribution is uniform within the cone. For each $\Delta\Omega$ the adequate
mean time between scatterings was chosen to keep the same magnetic field perturbations pattern.

The pattern is characterised by a value of the parameter $Q$ defined in section \ref{sec:Cracow}.
In this section we use more convenient parameter to work with. We have defined $C=-\log_{10} Q$.
If $C$ determines electromagnetic field perturbations pattern upstream of an ultrarelativistic
shock moving with a constant velocity, then with growing $C$ particles are able to accelerate
to higher energies in Cracow acceleration.

Particle momenta are measured in the unit of any momentum $p_0$. We have applied seed particles
momenta equal to $0.1$ because of computational reasons, so that the value $\log_{10}(p)=1$ means
that the seed particle momentum was increased hundred times.

The value of the mean magnetic field taken in the upstream plasma rest frame, $B_u$, is the unit
of the magnetic field. The particle charge, $e_0$, is the charge unit, so that the unit of time
is $p_0/(B_u e_0 c)$.

Our simulations are intended to model decelerating ultrarelativistic shocks,
so that we have chosen the equation of motion of the shock in the form

\[x_s(t)=-a\ln\frac{a+t}{a}\, ,\]
where $a>0$ is a constant and $x_s, t$ are the distance and the time measured
in the upstream plasma rest frame, respectively.

There is a lack of knowledge of the plasma conditions downstream of the decelerating
relativistic shock so that we have defined them. We have divided the motion of the shock into
discrete instants. We have applied a constant shock velocity at each instant. The downstream
plasma velocity and magnetic field are derived from the constant velocity relativistic shocks
(\citet{Bednarz96}) with the compression $R_h=3U_s^2$ for the hot plasma, where $U_s$ is the shock
speed in the upstream plasma rest frame. The shock is perpendicular, so that the instant
formulae are

\[U=\frac{3U_s^2-1}{2U_s}\, ,\;\;\;\; B=-B_uU_s\sqrt{\frac{9U_s^2-1}{1-U_s^2}}\, ,\]
where $U$ is the downstream plasma velocity close to the shock measured in the
upstream plasma rest frame and $B$ is the mean magnetic field close to the shock taken
in the downstream plasma rest frame.

We had to accept a rule in order to get $U(x,t)$ and $B(x,t)$ fields downstream
of the shock. The rule is that the decelerating shock 'puts the instant fields back'
and goes on. The shock sends the information about its present velocity into DOWN.
Plasma DOWN changes its magnetic and velocity fields according to the received information.
For simplicity the speed of the information is always equal to the speed of light.
The rule yields the fields

\begin{equation}
U(x,t)=\frac{3U_s^2(t-\Delta t)-1}{2U_s(t-\Delta t)},\label{eq:Ufield}
\end{equation}
\begin{equation}
B(x,t)=-U_s(t-\Delta t)\sqrt{\frac{9U_s^2(t-\Delta t)-1}{1-U_s^2(t-\Delta t)}},\label{eq:Bfield}
\end{equation}
where $\Delta t$ is derived from $\Delta t=x_s(t-\Delta t)-x$ and $x$, $t$ are the distance and
the time respectively, $x_s$ is the shock position and all quantities apart from $B(x,t)$
are measured in the upstream plasma rest frame. The above formulae are valid for $x_s(t)<x<t$.
An accelerating particle can not reach $x>t$.

We were injecting mono-energetic seed particles upstream of the shock with momenta distributed
uniformly on the sphere. The seed particle density along the shock path was constant.
The trajectories upstream of the shock were computed in the plasma rest frame also. Each
particle trajectory was followed using numerical computations until the particle escaped through
the free escape boundary placed far downstream from the shock or it reached the time larger than
the assumed upper limit. When the particle was wandering downstream of the shock its momentum
was Lorentz transformed to the respective plasma rest frame at each small step and its
trajectory was computed in the frame.

The initial and final Lorentz factors of the shock were chosen. The factors and the equation of
motion put the upper limit of time. We have divided the wandering time, measured in the upstream
plasma rest frame, into ten equal periods. The successive time bins are labelled numbers from
$0$ to $9$. The Lorentz factor range is labelled $\gamma_d=\gamma_i\_\gamma_f$, where
$\gamma_i$, $\gamma_f$ are the initial and final Lorentz factors, respectively. For example,
$\gamma_d=40\_5$ means that we started collecting data at the Lorentz factor of the shock
equal to $40$ and stopped at $5$.

\begin{figure}
\centering{\includegraphics[width=78mm]{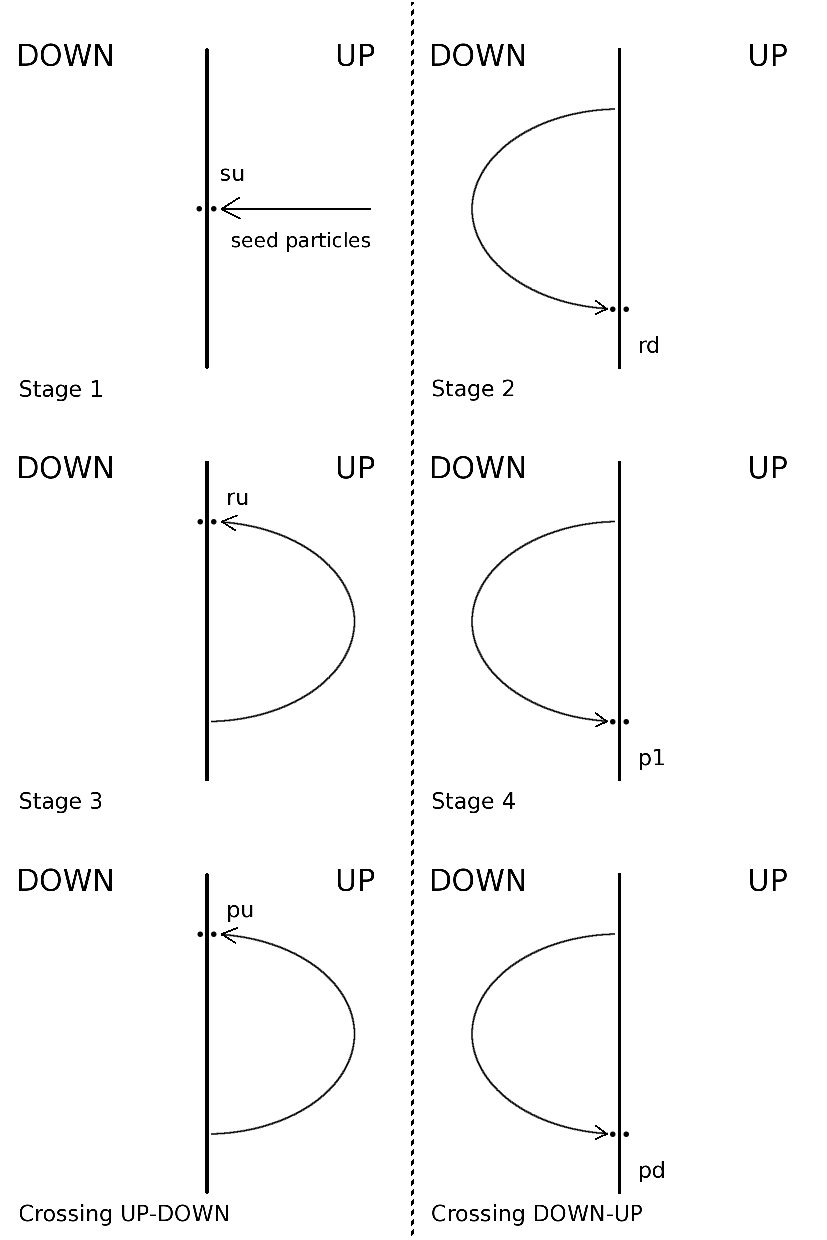}}
%\begin{center}%figure_4_1
%\FigureFile(78mm,){all_stages.eps}
%\end{center}%figure_08
\caption{Stages of particle crossings of the shock, where DOWN is downstream and UP is upstream
         of the shock.}
\label{fig:all_stages}
\end{figure}

The particle crossings of the shock were divided into stages as described in Fig.
\ref{fig:all_stages}. The stages are labelled 'su', 'rd', 'ru', 'p1', 'pu', 'pd'. The spectrum
of the particles crossing the shock from DOWN to UP for the first time is labelled 'rd', for
the second 'p1', for the third and following times 'pd'. The total spectrum of the particles
crossing the shock from DOWN to UP is the sum of 'rd', 'p1', 'pd' spectra and is labelled 'ad'.
We introduced the token $sp$ also, for instance $sp$='rd'.

Each time a particle crossed the shock the respective contribution was added to the given
momentum bin for the appropriate time bin and stage. The momentum and time are measured
in the upstream plasma rest frame.

We assume that particles emit synchrotron radiation. We neglect the self-absorption
and take into account the total emission coefficient only (\citet{Pacholczyk70})

\[\mathcal{E}_\nu=c_3 H \sin\vartheta \int_0^\infty N(E)F(x)dE,\]
where $c_3$ is a constant, $H$ magnetic field intensity, $\vartheta$ the angle between
the mean magnetic field and the direction toward the observer, $N(E)$ - the particle energy
distribution, $F(x)=x\int_x^\infty K_{5/3}(z)dz$, $K_{5/3}(z)$ is the Bessel function of
the second kind, $x=\nu/\nu_c$, $\nu$ - photon frequency, $\nu_c=c_1H\sin\vartheta E^2$,
$c_1$ a constant and $E$ is the particle energy.

We have calculated the spectral distribution of synchrotron radiation of accelerated
particles using the formula

\[I(\nu)=\int_0^\infty N(p) F(\frac{\nu}{p^2})dp,\]
where $I(\nu)$ is intensity of radiation and $N(p)$ the particle momentum distribution.
Units of $I(\nu)$ and $\nu$ are unimportant here and we applied a constant equal to 1.

In order to represent each set of spectra concisely we have introduced the vector
($\gamma_d$, $a$, $C$, $sp$). For instance, (20\_5, $10^2$, 3, pd) represents ten spectra
of particles produced by a relativistic shock slowing down from Lorentz factor 20 to 5
with the parameter of the equation of motion equal to $10^2$, the upstream disturbances pattern
measured by $C=3$ and particles crossing the shock from DOWN to UP for the third and following
times.

Fluctuations of the magnetic field downstream of the shock are unimportant now. They should be
strong enough to allow for effective reflections of seed particles. We have applied there
$C_d=-\log_{10} 50$, where $C_d$ is the value of $C$  downstream of the shock. The parameter $C_d$
is a variable and $C$ a constant in section \ref{sec:Fermi}.

\section{Slowing down shocks}\label{sec:slowing_down}
The velocity and magnetic fields ((\ref{eq:Ufield}), (\ref{eq:Bfield})) downstream of
the shock change the energy of wandering particles, so that they give some contribution to
the total acceleration. We did not analyse the amount of the contribution, but it is of little
significance to the paper. The presented figures suggest that the contribution does not influence
reflection and Cracow acceleration processes significantly, but one should remember that
for strong deceleration they are affected by the pattern of the velocity
and magnetic field downstream of the shock.

\begin{figure}
\centering{\includegraphics[width=78mm]{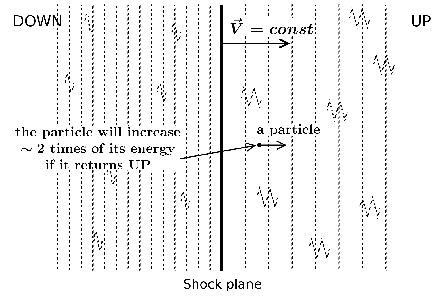}}
%\begin{center}%figure_5_1
%\FigureFile(78mm,){reason_2a.eps}
%\end{center}%figure_09
\caption{In constant velocity ultrarelativistic shocks particles double their energy along
         every lap.}
\label{fig:reason_2a}
\end{figure}
\begin{figure}
\centering{\includegraphics[width=78mm]{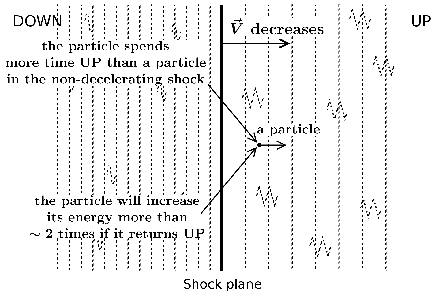}}
%\begin{center}%figure_5_2
%\FigureFile(78mm,){reason_2b.eps}
%\end{center}%figure_10
\caption{In slowing down ultrarelativistic shocks particles spend more time upstream of the shock
         and receive larger momentum gains.}
\label{fig:reason_2b}
\end{figure}

Cracow acceleration is very sensitive to the physical conditions upstream of the shock.
Its action in shocks moving with the constant velocity is very well known. The particles double
their energy along every lap there (Fig. \ref{fig:reason_2a}) and can not be accelerated
if there is a lack of strong disturbances of the particle movement upstream of the shock.

In slowing down shocks other factors give their contribution, but it is still the same
mechanism since the acceleration depends mainly on collective processes upstream of the shock.

Particles spend a bit more time (in units of the particle gyroradius) upstream of
slowing down than upstream of constant velocity shocks. This effect has two consequences. 

First, particles UP have smaller anisotropy in slowing down shocks. That means that the particles
receive larger energy gains along every lap (Fig. \ref{fig:reason_2b}) but their probability
of returning to the shock is smaller and in result they show steeper spectra and shifted
to a higher energy. This effect is valid for particles that enter UP with momenta more parallel
to the shock normal. Such particles increase their momentum more than others and therefore
the steeper and shifted spectra are seen at higher momenta.

Second, some particles enter UP with the directions of the momentum close to the limit of
the transition from DOWN to UP. Their interaction with upstream plasma is weak and their increase
of the momentum is small. If the shock has a constant velocity then a fraction of the particles
can not reach the shock again, but when the shock is slowing down the particles get an additional
time and change their trajectories in such a way that some particles of the fraction can reach
the shock again. This leads to flatter spectra at lower energies.

In result, Cracow acceleration in slowing down shocks produces flatter spectra at lower energies
and steeper at high energies.

There is still an issue to be addressed.  Particles are staying longer upstream of the shock
in an absolute unit of time (Fig. \ref{fig:reason_2b}). It means that particles that have higher
energies return to the shock later. This entails that the delaying particles are stopping the high
energy part of the spectrum from moving back to lower energies, what means that the part is moving
back slower or standing or moving ahead. We call the phenomenon lingering.

The stronger deceleration of the shock the shorter duration of the episode if $\gamma_i$ and
$\gamma_f$ are fixed. It means that the total time of the episode could be very small what
yields two effects.

The first effect refers to reflection process. Reflecting particles gain higher energies
if they are spending more time downstream of the shock. At first to the slowing down shock
(from DOWN) arrive particles which increase the momenta less and then with better energy gains.
These all particles have entered DOWN at the same Lorentz factor of the shock. If particles
with better energy gains arrive then they gain $\sim\gamma^2$ energy, but $\gamma$ is an average
of $\gamma$ at the crossing UP-DOWN and $\gamma$ at the crossing DOWN-UP. It means that
reflecting particles gain smaller maximum energies when $a$ is smaller. If $a$ is small and
particles that increase their momenta less cross the shock UP-DOWN at small $\gamma$ then they
can not return to the shock before $\gamma$ reaches $\gamma_f$. Instead, the particles with better
energy gains and larger $\gamma$ at the crossing UP-DOWN arrive. It means that reflecting particles
gain larger minimal energies when $a$ is smaller.

The second effect refers to Cracow acceleration. Particles accelerating at strong decelerating
shocks receive smaller energy gains because they have not enough time to receive larger.

\begin{table}
\caption{$\gamma_d$ - 10\_5, spectrum - $pd$}\label{tab:pd_10_5}
\begin{center}%table
\begin{tabular}{cccrcrl}
\hline
$a$ & $C$ & $n_p$ & $\nu_p$ & $n_n$ & $\nu_n$ & lag \\
\hline
$10^{2}$ & 4 & \nodata & \nodata & \nodata & \nodata & R  \\
$10^{3}$ & 2 & \nodata & \nodata & 2 & 2.67 & N$\uparrow$ \\
$10^{3}$ & 3 & \nodata & \nodata & 2 & 2.44 & N$\uparrow$ \\
$10^{3}$ & 4 & \nodata & \nodata & 2 & 2.28 & N$\uparrow$ \\
$10^{4}$ & 2 & \nodata & \nodata & 1 & 4.26 & N$\uparrow$ \\
$10^{4}$ & 3 & \nodata & \nodata & 1 & 4.06 & N$\uparrow$ \\
$10^{4}$ & 4 & \nodata & \nodata & 1 & 3.80 & N$\uparrow$ \\
$10^{5}$ & 2 & \nodata & \nodata & \nodata & \nodata & D  \\
\hline
\end{tabular}
\end{center}%table
\end{table}

\begin{table}
\caption{$\gamma_d$ - 10\_5, spectrum - $p1$}\label{tab:p1_10_5}
\begin{center}%table
\begin{tabular}{cccrcrl}
\hline
$a$ & $C$ & $n_p$ & $\nu_p$ & $n_n$ & $\nu_n$ & lag \\
\hline
$10^{2}$ & 4 & \nodata & \nodata & \nodata & \nodata & R  \\
$10^{3}$ & 2 & \nodata & \nodata & 2 & 3.78 & N$\uparrow$ \\
$10^{3}$ & 3 & \nodata & \nodata & 2 & 4.03 & N$\uparrow$ \\
$10^{3}$ & 4 & \nodata & \nodata & 1 & 3.52 & N$\uparrow$ \\
$10^{4}$ & 2 & \nodata & \nodata & \nodata & \nodata & D  \\
\hline
\end{tabular}
\end{center}%table
\end{table}

\begin{table}
\caption{$\gamma_d$ - 10\_5, spectrum - $ad$}\label{tab:ad_10_5}
\begin{center}%table
\begin{tabular}{cccrcrl}
\hline
$a$ & $C$ & $n_p$ & $\nu_p$ & $n_n$ & $\nu_n$ & lag \\
\hline
60       & 2 & \nodata & \nodata & 8 & 1.98 & N \\
60       & 3 & \nodata & \nodata & 8 & 1.95 & N \\
60       & 4 & \nodata & \nodata & 8 & 1.90 & N \\
$10^{2}$ & 2 & \nodata & \nodata & 6 & 2.06 & N$\uparrow$ \\
$10^{2}$ & 3 & \nodata & \nodata & 6 & 1.92 & N$\uparrow$ \\
$10^{2}$ & 4 & \nodata & \nodata & 6 & 1.94 & N$\uparrow$ \\
$10^{3}$ & 2 & \nodata & \nodata & 1 & 2.25 & N$\uparrow$ \\
$10^{3}$ & 3 & \nodata & \nodata & 1 & 2.25 & N$\uparrow$ \\
$10^{3}$ & 4 & \nodata & \nodata & 1 & 2.16 & N$\uparrow$ \\
$10^{4}$ & 2 & \nodata & \nodata & 0 & 2.74 & N$\uparrow$ \\
$10^{4}$ & 3 & \nodata & \nodata & 0 & 2.52 & N$\uparrow$ \\
$10^{4}$ & 4 & \nodata & \nodata & 0 & 2.38 & N$\uparrow$ \\
$10^{5}$ & 2 & \nodata & \nodata & \nodata & \nodata & D  \\
\hline
\end{tabular}
\end{center}%table
\end{table}

\begin{table}
\caption{$\gamma_d$ - 10\_5, spectrum - $rd$}\label{tab:rd_10_5}
\begin{center}%table
\begin{tabular}{cccrcrl}
\hline
$a$ & $C$ & $n_p$ & $\nu_p$ & $n_n$ & $\nu_n$ & lag \\
\hline
60       & 2 & 3 & 3.16 & \nodata & \nodata & P$\downarrow$ \\
60       & 3 & 3 & 2.99 & \nodata & \nodata & P$\downarrow$ \\
60       & 4 & 3 & 3.23 & \nodata & \nodata & P$\downarrow$ \\
$10^{2}$ & 2 & 2 & 2.87 & \nodata & \nodata & P$\downarrow$ \\
$10^{2}$ & 3 & 2 & 2.94 & \nodata & \nodata & P$\downarrow$ \\
$10^{2}$ & 4 & 2 & 2.90 & \nodata & \nodata & P$\downarrow$ \\
$10^{3}$ & 2 & 0 & 2.47 & \nodata & \nodata & P \\
$10^{3}$ & 3 & 0 & 2.54 & \nodata & \nodata & P \\
$10^{3}$ & 4 & 0 & 2.49 & \nodata & \nodata & P \\
$10^{4}$ & 2 & \nodata & \nodata & \nodata & \nodata & D \\
\hline
\end{tabular}
\end{center}%table
\end{table}

\section{Results}\label{sec:results}
From graphs of particle spectra one can get the energy spectral index

\[\beta=1-\frac{\Delta[\log_{10}(dN)/d\log_{10}(p)]}{\Delta[\log_{10}(p)]}\, .\]

In each of photon spectra we distinguish the frequency $\nu_{max}$ where the maximum intensity
$I(\nu_{max})$ is reached. In principle, we limit ourselves to frequencies larger
than $\nu_{max}$.

The first lap in the process of Cracow acceleration is represented by 'p1' spectra. They are
similar to corresponding 'pd' spectra. We have not presented a 'p1' spectrum, but the observation
is obvious as one compares Tables \ref{tab:pd_10_5}, \ref{tab:p1_10_5} and \ref{tab:rd_10_5}.
Our first result is that reflection process does not influence Cracow acceleration from
the first lap and therefore we can treat these processes as separate.

\begin{figure}
\centering{\includegraphics[width=78mm]{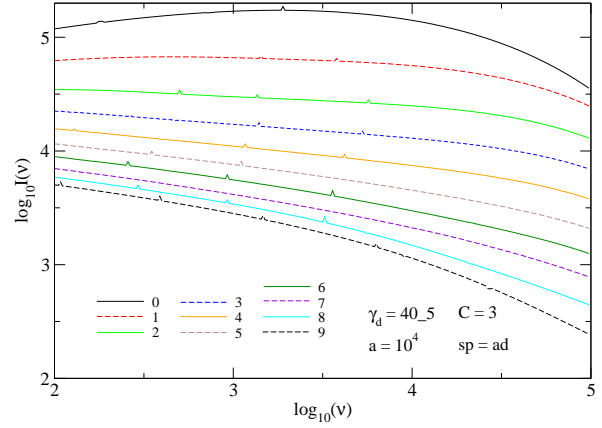}}
%\begin{center}%figure_6_01
%\FigureFile(78mm,){ds_all_ad.eps}
%\end{center}%figure_11
\caption{Photon spectra of (40\_5, $10^4$, 3, ad).}
\label{fig:ds_all_ad}
\end{figure}

\begin{figure}
\centering{\includegraphics[width=78mm]{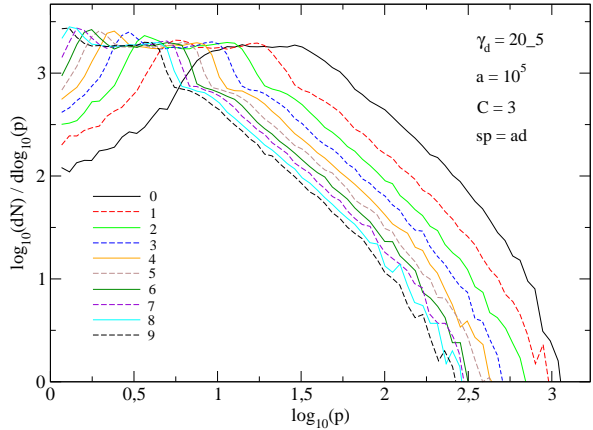}}
%\begin{center}%figure_6_02
%\FigureFile(78mm,){v_const_ad.eps}
%\end{center}%figure_12
\caption{Particle spectra of (20\_5, $10^5$, 3, ad). The flat part of the spectra has got
         $\beta$ in the range $(2.4, 2.5)$.}
\label{fig:v_const_ad}
\end{figure}

In Fig. \ref{fig:ds_all_ad}, \ref{fig:v_const_ad}, \ref{fig:v_const_rd} and
\ref{fig:v_const_pd} are presented spectra that are produced by practically constant velocity
shocks. The flat part of 'pd' spectra (Fig. \ref{fig:v_const_pd}) has got $\beta$ in the range
2.5-2.7 what is larger than the limiting value of $\beta$ produced by Cracow acceleration
in constant velocity shocks.

\begin{figure}
\centering{\includegraphics[width=78mm]{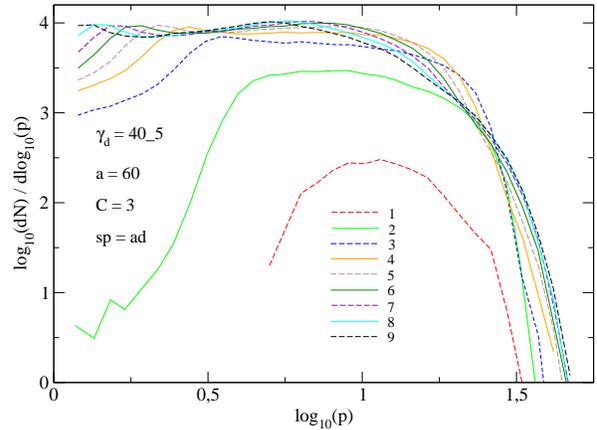}}
%\begin{center}%figure_6_05
%\FigureFile(78mm,){v_decel_ad.eps}
%\end{center}%figure_15
\caption{Particle spectra of (40\_5, 60, 3, ad). Maximum momenta are small because particles
         have not enough time to get larger. The value of $\beta$ of the flat part of the spectrum
         at the time 2 is in the range $1.2-2.2$, at the time 4 is equal to 1.0 and
         at the time 7 is equal to 0.7.}
\label{fig:v_decel_ad}
\end{figure}

\begin{figure}
\centering{\includegraphics[width=78mm]{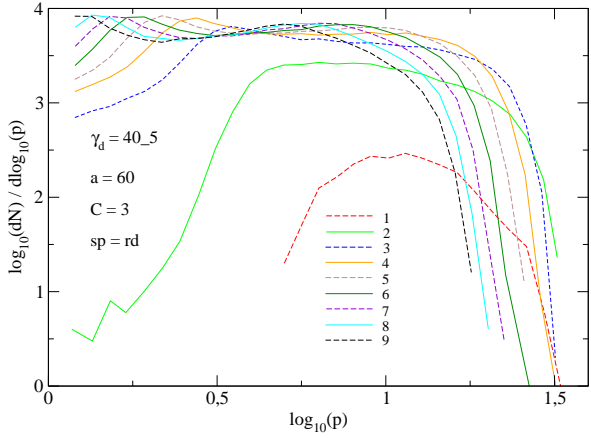}}
%\begin{center}%figure_6_06
%\FigureFile(78mm,){v_decel_rd.eps}
%\end{center}%figure_16
\caption{Particle spectra of (40\_5, 60, 3, rd). The smaller range of maximum energies
         of particles reflecting from the strong decelerating shock is visible.}
\label{fig:v_decel_rd}
\end{figure}

\begin{figure}
\centering{\includegraphics[width=78mm]{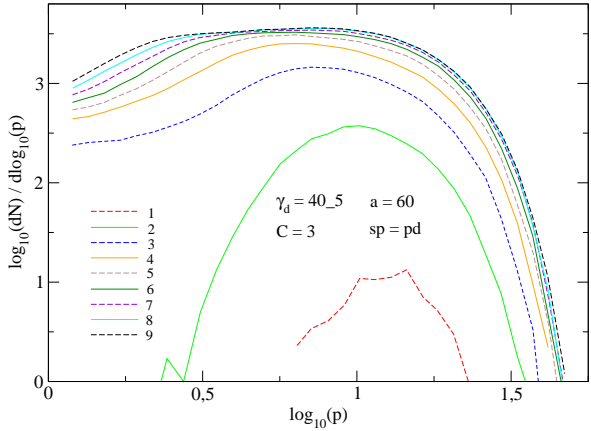}}
%\begin{center}%figure_6_07
%\FigureFile(78mm,){v_decel_pd.eps}
%\end{center}%figure_17
\caption{Particle spectra of (40\_5, 60, 3, pd). Cracow acceleration process in slowing down
         shocks produces flatter spectra at lower energies and steeper at high energies.}
\label{fig:v_decel_pd}
\end{figure}

In next three figures (Fig. \ref{fig:v_decel_ad}, \ref{fig:v_decel_rd}, \ref{fig:v_decel_pd})
the particle spectra produced by a strong decelerating shock are shown. If they are compared
with the three preceding figures then one can see to what extent some remarks presented in
the previous section are correct.

Our primary task is to present how lags are formed in photon spectra of slowing down
ultrarelativistic shocks. We have examined the spectra and collected results in tables.
The description of each table contains the values of $\gamma_d$, $sp$ and two first
columns the values of $a$, $C$.

A lag will be visible as an intersection of two consecutive photon spectra (produced in two
consecutive periods of time). Let as denote the frequency at which the intersection occurs as
$\nu_i$, the number of the first spectrum as $n$ and the second as $n+1$. If photon intensity of
the $n+1$ spectrum is higher than the intensity of the $n$ spectrum just before $\nu_i$ (at lower
frequency), then the positive lag appears. If the intensity of the $n+1$ spectrum is lower than
the intensity of the previous spectrum just before $\nu_i$, then the negative lag appears.

The first occurrence of the positive lag in the set of photon spectra for consecutive periods of
time was put into columns third and fourth, where $n_p$ is the number of the previous spectrum
and $\nu_p$ is the frequency at which the intersection occurs. The first occurrence of
the negative lag was collected in the fifth and sixth columns, where $n_n$ is the number of
the previous spectrum and $\nu_n$ is the frequency at which the intersection occurs.

The seventh column is marked by 'lag' and contains the information how lags in the photon
spectra are changing. The photon spectra that are only increasing (decreasing) their intensities
at all frequencies are denoted as R (D). If the intensities at first are increasing and next
they are decreasing at all frequencies, then they are denoted as RD.

\begin{table}
\caption{$\gamma_d$ - 40\_5, spectrum - $rd$}\label{tab:rd_40_5}
\begin{center}%table
\begin{tabular}{cccrcrl}
\hline
$a$ & $C$ & $n_p$ & $\nu_p$ & $n_n$ & $\nu_n$ & lag \\
\hline
60       & 2 & 2 & 3.85 & \nodata & \nodata & P$\downarrow$ \\
60       & 3 & 2 & 3.67 & \nodata & \nodata & P$\downarrow$ \\
60       & 4 & 2 & 3.55 & \nodata & \nodata & P$\downarrow$ \\
$10^{2}$ & 2 & 1 & 4.08 & \nodata & \nodata & P$\downarrow$ \\
$10^{2}$ & 3 & 1 & 3.94 & \nodata & \nodata & P$\downarrow$ \\
$10^{2}$ & 4 & 1 & 4.17 & \nodata & \nodata & P$\downarrow$ \\
$10^{3}$ & 2 & \nodata & \nodata & \nodata & \nodata & D    \\
\hline
\end{tabular}
\end{center}%table
\end{table}

\begin{table}
\caption{$\gamma_d$ - 40\_5, spectrum - $pd$}\label{tab:pd_40_5}
\begin{center}%table
\begin{tabular}{cccrcrl}
\hline
$a$ & $C$ & $n_p$ & $\nu_p$ & $n_n$ & $\nu_n$ & lag \\
\hline
$10^{2}$ & 4 & \nodata & \nodata & \nodata & \nodata & R  \\
$10^{3}$ & 2 & \nodata & \nodata & \nodata & \nodata & RD \\
$10^{3}$ & 3 & \nodata & \nodata & \nodata & \nodata & RD \\
$10^{3}$ & 4 & \nodata & \nodata & 1 & 4.23 & N$\uparrow$ \\
$10^{4}$ & 2 & \nodata & \nodata & \nodata & \nodata & D  \\
\hline
\end{tabular}
\end{center}%table
\end{table}

\begin{table}
\caption{$\gamma_d$ - 20\_5, spectrum - $ad$}\label{tab:ad_20_5}
\begin{center}%table
\begin{tabular}{cccrcrl}
\hline
$a$ & $C$ & $n_p$ & $\nu_p$ & $n_n$ & $\nu_n$ & lag \\
\hline
60       & 1 & 3 & 2.87 & 6 & 3.45 & P$\downarrow$N$\downarrow$ \\
60       & 2 & 3 & 2.89 & 3 & 3.63 & P$\downarrow$N$\downarrow$ \\
60       & 3 & 5 & 1.56 & 5 & 2.78 & PN$\downarrow$ \\
60       & 4 & 5 & 1.45 & 5 & 2.66 & PN$\downarrow$ \\
$10^{2}$ & 1 & 2 & 2.76 & 3 & 4.01 & P$\downarrow$N$\downarrow$ \\
$10^{2}$ & 2 & 2 & 2.99 & 2 & 3.80 & P$\downarrow$N$\downarrow$ \\
$10^{2}$ & 3 & 3 & 2.03 & 3 & 3.06 & PN$\downarrow$ \\
$10^{2}$ & 4 & \nodata & \nodata & 4 & 2.83 & N$\downarrow$ \\
$10^{3}$ & 1 & \nodata & \nodata & 1 & 3.87 & N$\uparrow$ \\
$10^{3}$ & 2 & \nodata & \nodata & 1 & 3.66 & N$\uparrow$ \\
$10^{3}$ & 3 & \nodata & \nodata & 1 & 3.56 & N$\uparrow$ \\
$10^{3}$ & 4 & \nodata & \nodata & 1 & 3.47 & N$\uparrow$ \\
$10^{4}$ & 1 & \nodata & \nodata & 0 & 4.52 & N \\
$10^{4}$ & 2 & \nodata & \nodata & 0 & 4.47 & N \\
$10^{4}$ & 3 & \nodata & \nodata & 0 & 4.31 & N \\
$10^{4}$ & 4 & \nodata & \nodata & 0 & 4.14 & N \\
$10^{5}$ & 1 & \nodata & \nodata & \nodata & \nodata & D \\
\hline
\end{tabular}
\end{center}%table
\end{table}

\begin{table}
\caption{$\gamma_d$ - 40\_5, spectrum - $ad$}\label{tab:ad_40_5}
\begin{center}%table
\begin{tabular}{cccrcrl}
\hline
$a$ & $C$ & $n_p$ & $\nu_p$ & $n_n$ & $\nu_n$ & lag \\
\hline
60       & 2 & 2 & 3.90 & 3 & 4.13 & P$\downarrow$N$\downarrow$ \\
60       & 3 & 3 & 2.86 & 3 & 3.64 & P$\downarrow$N$\downarrow$ \\
60       & 4 & 4 & 2.10 & 4 & 3.13 & PN$\downarrow$ \\
$10^{2}$ & 2 & 2 & 2.65 & 3 & 4.13 & PN$\downarrow$ \\
$10^{2}$ & 3 & 2 & 2.80 & 2 & 4.03 & PN$\downarrow$ \\
$10^{2}$ & 4 & 5 & 4.48 & \nodata & \nodata & P$\downarrow$ \\
$10^{3}$ & 2 & 0 & 2.92 & \nodata & \nodata & P \\
$10^{3}$ & 3 & 0 & 3.50 & 0 & 4.21 & PN$\uparrow$ \\
$10^{3}$ & 4 & \nodata & \nodata & 1 & 4.42 & N$\uparrow$ \\
$10^{4}$ & 2 & \nodata & \nodata & \nodata & \nodata & D  \\
\hline
\end{tabular}
\end{center}%table
\end{table}

\begin{table}
\caption{$\gamma_d$ - 20\_10, spectrum - $ad$}\label{tab:ad_20_10}
\begin{center}%table
\begin{tabular}{cccrcrl}
\hline
$a$ & $C$ & $n_p$ & $\nu_p$ & $n_n$ & $\nu_n$ & lag \\
\hline
$10^{3}$ & 2 & \nodata & \nodata & 4 & 3.49 & NN$\uparrow$ \\
$10^{3}$ & 4 & \nodata & \nodata & 5 & 3.26 & N$\uparrow$  \\
\hline
\end{tabular}
\end{center}%table
\end{table}

\begin{table}
\caption{$\gamma_d$ - 40\_10, spectrum - $ad$}\label{tab:ad_40_10}
\begin{center}%table
\begin{tabular}{cccrcrl}
\hline
$a$ & $C$ & $n_p$ & $\nu_p$ & $n_n$ & $\nu_n$ & lag \\
\hline
$10^{3}$ & 2 & 2 & 2.81 & 4 & 4.91 & PN$\downarrow$ \\
$10^{3}$ & 4 & 2 & 3.54 & 2 & 4.56 & PN$\downarrow$ \\
\hline
\end{tabular}
\end{center}%table
\end{table}

If R appears at some values of $a$ and $C$ then for all smaller values of $a$ and for the same
value of $a$ and smaller values of $C$ R appears also. In that case subsequent items of R
are not included in tables. If D appears at some values of $a$ and $C$ then for all larger
values of $a$ and for the same value of $a$ and larger values of $C$ D appears also.
In that case subsequent items of D are not included in tables.

If $n$ spectrum and $n+1$ spectrum build a positive lag at frequency $\nu_p$ and then $n+1$
spectrum and $n+2$ spectrum build a positive lag at frequency $\nu_{p2}<\nu_p$ and so on
(or not), then the spectra are denoted as P$\downarrow$. If only one positive lag appears then
the spectra are denoted as P. Here it is necessary to add that we did not take into account
positive lags that appeared at frequencies lower than $\nu_{max}$. In fact positive lags are
always P$\downarrow$ and are moving very fast towards low frequencies passing through
$\nu_{max}$.

If $n$ spectrum and $n+1$ spectrum build a negative lag at frequency $\nu_n$ and then $n+1$
spectrum and $n+2$ spectrum build a negative lag at frequency $\nu_{n2}<\nu_n$ and so on
(or not), then the spectra are denoted as N$\downarrow$. If only one negative lag appears then
the spectra are denoted as N. This negative lag is moving slowly towards low frequencies and
stops at a high frequency far before $\nu_{max}$.

\begin{figure}
\centering{\includegraphics[width=78mm]{figure_6_03.eps}}
%\begin{center}%figure_6_03
%\FigureFile(78mm,){v_const_rd.eps}
%\end{center}%figure_13
\caption{Particle spectra of (20\_5, $10^5$, 3, rd).}
\label{fig:v_const_rd}
\end{figure}
\begin{figure}
\centering{\includegraphics[width=78mm]{figure_6_04.eps}}
%\begin{center}%figure_6_04
%\FigureFile(78mm,){v_const_pd.eps}
%\end{center}%figure_14
\caption{Particle spectra of (20\_5, $10^5$, 3, pd). The flat part of the spectra has got
         $\beta$ in the range $(2.5, 2.7)$.}
\label{fig:v_const_pd}
\end{figure}

If $n$ spectrum and $n+1$ spectrum build a negative lag at frequency $\nu_n$ and then $n+1$
spectrum and $n+2$ spectrum build a negative lag at frequency $\nu_{n2}>\nu_n$ and so on
(or not), then the spectra are denoted as N$\uparrow$. If only one negative lag appears then
the spectra are denoted as N. These types of spectra are always N$\uparrow$. The case of N
arises due to the limitation of data and large errors at high frequencies.

\begin{figure}
\centering{\includegraphics[width=78mm]{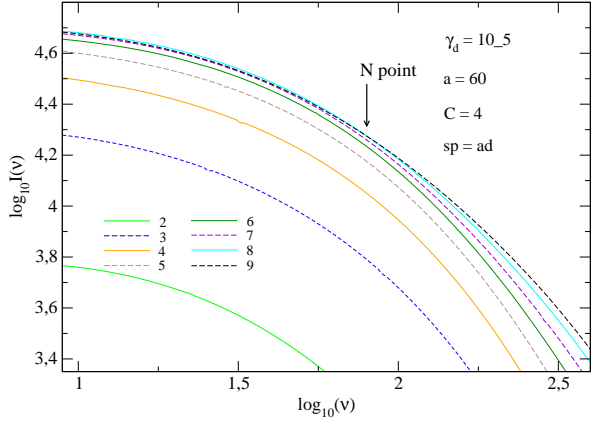}}
%\begin{center}%figure_6_08
%\FigureFile(78mm,){ns_all_ad.eps}
%\end{center}%figure_18
\caption{Photon spectra of (10\_5, 60, 4, ad). In real physical conditions the shock
         should produce the lag of the type N$\uparrow$.}
\label{fig:ns_all_ad}
\end{figure}

\begin{figure}
\centering{\includegraphics[width=78mm]{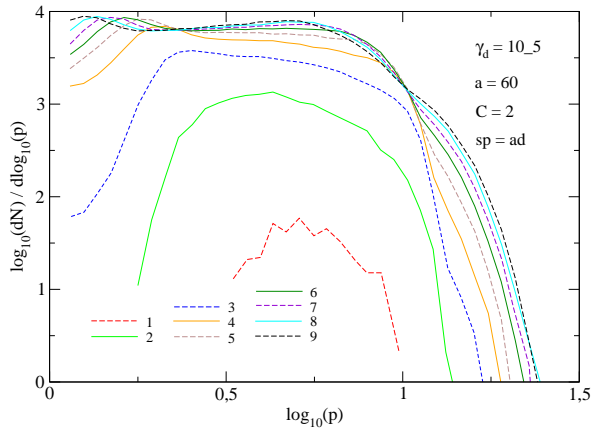}}
%\begin{center}%figure_6_09
%\FigureFile(78mm,){n_struct_ad.eps}
%\end{center}%figure_19
\caption{Particle spectra of (10\_5, 60, 2, ad). The value of $\beta$ of the flat part of
         the spectrum at the time 3 is equal to 1.7, at the time 4 is equal to 1.4 and
         at the time 7 is equal to 0.9.}
\label{fig:n_struct_ad}
\end{figure}

\begin{figure}
\centering{\includegraphics[width=78mm]{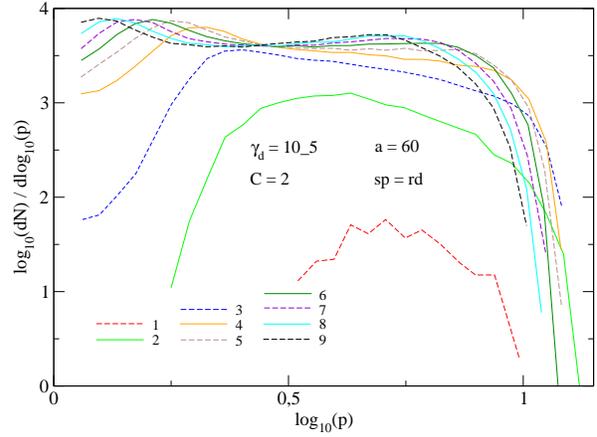}}
%\begin{center}%figure_6_10
%\FigureFile(78mm,){n_struct_rd.eps}
%\end{center}%figure_20
\caption{Particle spectra of (10\_5, 60, 2, rd). The high-energy tail moves back too slowly
         to be able to produce a positive lag in 'ad' spectra.}
\label{fig:n_struct_rd}
\end{figure}

An exception exists for small initial Lorentz factors of the shock and small $a$ (Fig.
\ref{fig:ns_all_ad}, \ref{fig:n_struct_ad}, \ref{fig:n_struct_rd}). In Table \ref{tab:ad_10_5}
for the case of (10\_5, 60, any, ad) the spectra number 8 and 9 build the negative lag N
(Fig. \ref{fig:ns_all_ad}). First, we do not expect such small value of $a$ for small initial
Lorentz factors. Second, $a$ must grow since a moment. We expect that in real physical
conditions such shocks form spectra of the type N$\uparrow$.

\begin{figure}
\centering{\includegraphics[width=78mm]{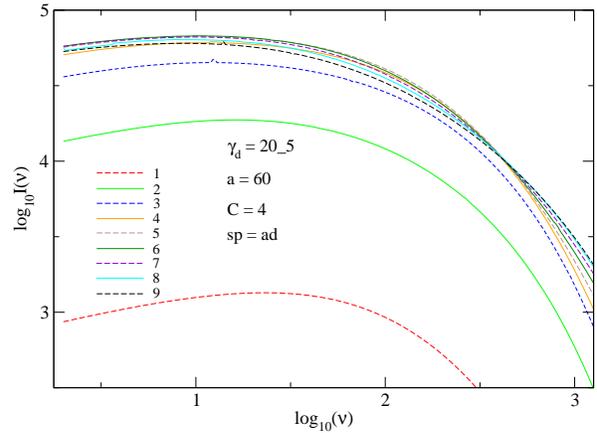}}
%\begin{center}%figure_6_11
%\FigureFile(78mm,){pn_all_ad.eps}
%\end{center}%figure_21
\caption{Photon spectra of (20\_5, 60, 4, ad). An example of the structure
         P$\downarrow$N$\downarrow$ without details.}
\label{fig:pn_all_ad}
\end{figure}

\begin{figure}
\centering{\includegraphics[width=78mm]{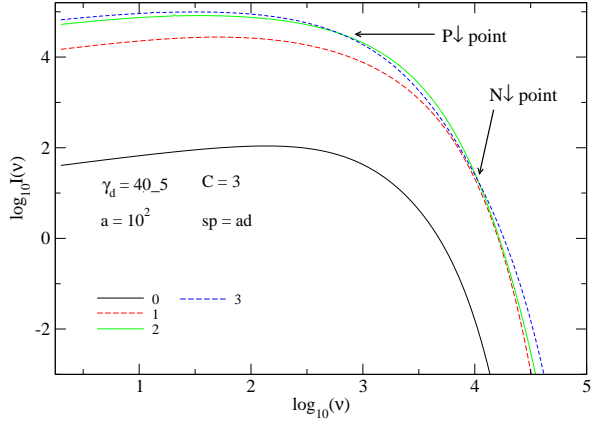}}
%\begin{center}%figure_6_12
%\FigureFile(78mm,){pn_struct_ad.eps}
%\end{center}%figure_22
\caption{Photon spectra of (40\_5, $10^2$, 3, ad). An example of the structure
         P$\downarrow$N$\downarrow$ with details.}
\label{fig:pn_struct_ad}
\end{figure}

The structure P$\downarrow$ often appears together with the structure N$\downarrow$, in this
case we will consider the structure P$\downarrow$N$\downarrow$. The untypical structure
PN$\uparrow$ in (40\_5, $10^3$, 3, ad) arose from the small time resolution. An example of
the structure P$\downarrow$N$\downarrow$ without details is presented in Fig.
\ref{fig:pn_all_ad} and with details in Fig. \ref{fig:pn_struct_ad}.

Data presented in tables with 'rd' (Tables \ref{tab:rd_10_5}, \ref{tab:rd_40_5}) and 'pd' (Tables
\ref{tab:pd_10_5}, \ref{tab:pd_40_5}) spectra give us the very important result that individual
processes give their own contribution to the general spectrum. Reflection process generates
the structure P$\downarrow$ (the positive lag) and Cracow acceleration the structure N$\uparrow$
(the negative lag). That is the only result we have obtained from the tables.

We have put only one item N$\uparrow$ into Table \ref{tab:pd_40_5} because of low accuracy.
However, we detected for $a=10^3$, $C=3$ the structure N at $n_n=1$ and $\nu_n=4.4$ that seems to
be N$\uparrow$ and for $a=10^2$, $C=4$ the structure N at $n_n=5$ and $\nu_n=4.6$. Moreover, 'p1'
spectra give their contribution. In 'p1' spectra we detected for $a=10^3$, $C=3$ the structure
at $n_n=1$ and $\nu_n=4.6$ that seems to be N$\downarrow$, but it could be N$\uparrow$. In spectra
(40\_5, $10^2$, any, p1) we detected positive and negative lags and in spectra (40\_5, 60, any, p1)
positive lags only. The spectra for $sp$='p1' are not pure and reflection process influences them
at strong deceleration and large $\gamma_i$. The low accuracy results from detection of lags
at high $\nu$ and too large time step between consecutive spectra. Henceforth, the tables with 'ad'
(Tables \ref{tab:ad_10_5}, \ref{tab:ad_20_5}, \ref{tab:ad_40_5}, \ref{tab:ad_20_10},
\ref{tab:ad_40_10}) spectra will be discussed.

\begin{figure}
\centering{\includegraphics[width=78mm]{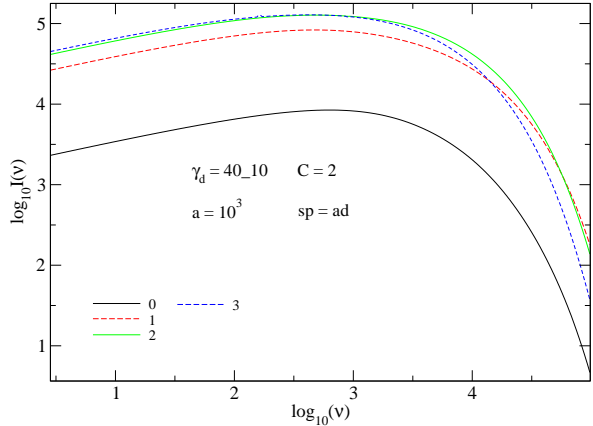}}
%\begin{center}%figure_6_13
%\FigureFile(78mm,){p_pure_ad.eps}
%\end{center}%figure_23
\caption{Photon spectra of (40\_10, $10^3$, 2, ad). The spectra are practically purely
         P$\downarrow$.}
\label{fig:p_pure_ad}
\end{figure}
\begin{figure}
\centering{\includegraphics[width=78mm]{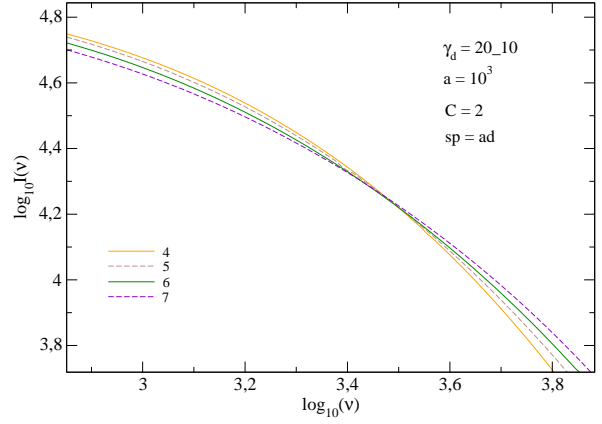}}
%\begin{center}%figure_6_14
%\FigureFile(78mm,){n_still_ad.eps}
%\end{center}%figure_24
\caption{Photon spectra of (20\_10, $10^3$, 2, ad). We have labelled the spectra NN$\uparrow$
         since the point N does not move at first.}
\label{fig:n_still_ad}
\end{figure}

When one is examining the tables, one can see that the structure P$\downarrow$N$\downarrow$
occurs at large initial Lorentz factors of the shock and strong deceleration. If $\gamma_i$
is smaller and/or $a$ is larger then the spectra exhibit the structure N$\uparrow$.

To obtain a better resolution of (40\_5, $10^3$, any, ad) and (20\_5, $10^3$, any, ad) spectra
we have performed the same simulations as previously but $\gamma_{f}$ was equal to 10 instead
of 5. The additional simulations were following (40\_10, $10^3$, 2, any),
(40\_10, $10^3$, 4, any), (20\_10, $10^3$, 2, any), (20\_10, $10^3$, 4, any) and the outcome
was put into Tables \ref{tab:ad_20_10} and \ref{tab:ad_40_10}.

The spectra of (40\_10, $10^3$, 2, ad) show the structure P$\downarrow$ at low frequencies and
the unimportant structure N$\downarrow$ at higher frequencies and later times what is
not visible for $\gamma_d=40\_5$. The spectra represent almost pure P$\downarrow$ that is
not suppressed by Cracow acceleration (Fig. \ref{fig:p_pure_ad}). We neglected results if $\nu$
was high, here we neglected P for $\nu=4.75$ and we have P instead of P$\downarrow$.

The case of (40\_5, $10^2$, 4, ad) is in fact P$\downarrow$N$\downarrow$, but N$\downarrow$
is outside the range of detectable results (Table \ref{tab:ad_40_5}).

\begin{figure}
\centering{\includegraphics[width=78mm]{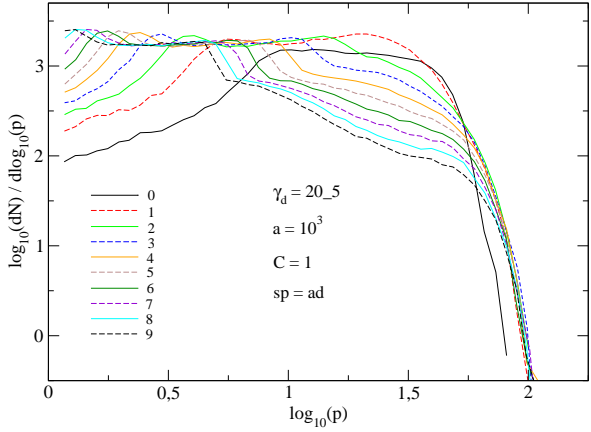}}
%\begin{center}%figure_6_15
%\FigureFile(78mm,){s_stop_ad.eps}
%\end{center}%figure_25
\caption{Particle spectra of (20\_5, $10^3$, 1, ad). The lingering of the high-energy tail that
         arises from Cracow acceleration operating at the slowing down shock is visible.
         The value of $\beta$ of the flat part of the spectrum at the time 0 is equal to 1.2
         and then there are two flat parts, at the time 2 $\beta=0.7$ and $\beta=2.1$,
         and at the time 5 $\beta=0.9$ and $\beta=1.8$.}
\label{fig:s_stop_ad}
\end{figure}
\begin{figure}
\centering{\includegraphics[width=78mm]{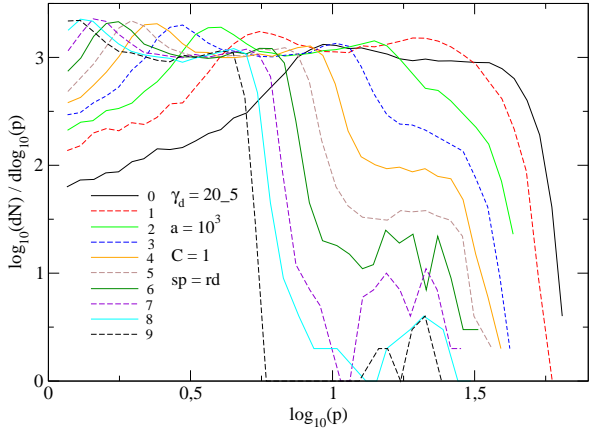}}
%\begin{center}%figure_6_16
%\FigureFile(78mm,){s_stop_rd.eps}
%\end{center}%figure_26
\caption{Particle spectra of (20\_5, $10^3$, 1, rd).}
\label{fig:s_stop_rd}
\end{figure}
\begin{figure}
\centering{\includegraphics[width=78mm]{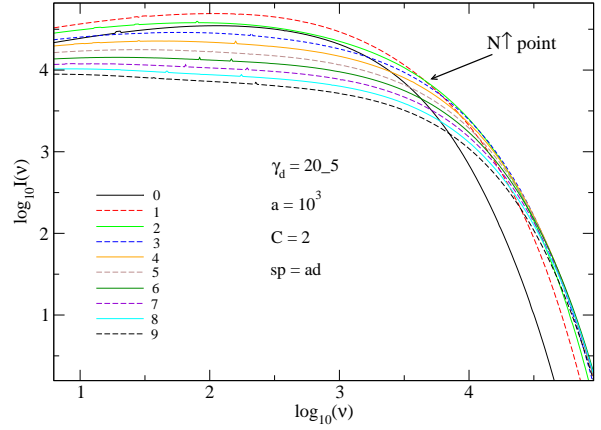}}
%\begin{center}%figure_6_17
%\FigureFile(78mm,){nu_all_ad.eps}
%\end{center}%figure_27
\caption{Photon spectra of (20\_5, $10^3$, 2, ad). An example of the structure N$\uparrow$.}
\label{fig:nu_all_ad}
\end{figure}

We have analysed the particle and photon spectra and reached the following conclusions,

{\itshape a) the structure P$\downarrow$ is formed when the sudden fall in the spectra of
reflected particles 'rd' is not suppressed by Cracow acceleration,}

{\itshape b) the structure P$\downarrow$N$\downarrow$ is formed when lingering of
the high-energy tail of Cracow acceleration spectra 'pd' is accompanied by the sudden fall in
the spectra of reflected particles 'rd',}

{\itshape c) the structure N$\uparrow$ is formed due to lingering of the high-energy tail
of Cracow acceleration spectra 'pd' in the presence of the weak fall in the spectra of reflected
particles 'rd'.}

The fall in the spectra of reflected particles is best visible when one is looking at
the high-energy tail of the spectra. Before doing so it is convenient to raise the particle
spectra of initial times to the density level of the remaining spectra. Then, the high-energy tail
will almost always move towards low energies.

We have done rough estimation of the fall and introduced the parameter: $-\Delta p/\Delta t$,
where $\Delta p$ is the difference of the momenta measured at a constant particle density
between the two particle spectra that differ by the time $\Delta t$. The value of the parameter
was measured at the high-energy tail. The results are collected in Table \ref{tab:rd_fall}.
In the table $n_1$, $n_2$ are numbers of spectra for times $t_1$, $t_2$, where momenta $p_1$, $p_2$
are measured at a selected density, respectively.

\begin{table}
\caption{Power of falling of 'rd' spectra}\label{tab:rd_fall}
\begin{center}%table
\begin{tabular}{cccrcr}
\hline
$\gamma_d$ & $a$ & $C$ & $n_1$ & $n_2$ & $\frac{p_1-p_2}{t_2-t_1}$ \\
\hline
40\_5  &     60 & 3 & 3 & 6 & 19.8  \\
10\_5  &     60 & 2 & 4 & 8 & 4.59  \\
40\_5  & $10^5$ & 3 & 0 & 1 & 0.412 \\
40\_5  & $10^5$ & 3 & 3 & 5 & 0.012 \\
20\_10 & $10^3$ & 2 & 3 & 8 & 7.83  \\
40\_10 & $10^3$ & 2 & 2 & 6 & 29.3  \\
\hline
\end{tabular}
\end{center}%table
\end{table}

In spectra of (20\_10, $10^3$, 2, ad) we have observed a weak structure P$\downarrow$ at
frequencies lower than $\nu_{max}$. We have estimated that the case is close to the boundary
between appearing and not-appearing of P$\downarrow$ in 'ad' spectra for $C\gtrsim 2$ and
$C\lesssim 3-4$. The boundary is at $-\Delta p/\Delta t\simeq8$. One should remember that both
the parameter and the measurement are rough.

We denoted the spectra of (20\_10, $10^3$, 2, ad) as NN$\uparrow$ since N keeps still at first
(Fig. \ref{fig:n_still_ad}).

The lingering of the high-energy tail in Cracow acceleration spectra is presented in Fig.
\ref{fig:s_stop_ad} and related 'rd' spectra in Fig. \ref{fig:s_stop_rd}. The tail moves ahead
if C is larger. Photon spectra with the structure N$\uparrow$ are presented in Fig.
\ref{fig:nu_all_ad}.

The phenomenon of the production of lags is simple. When one of the two acceleration processes
dominates the other then appears P$\downarrow$ or N$\uparrow$. When they compete then P$\downarrow$
is pulling the progenitor N$\uparrow$ and changes it into N$\downarrow$. This interaction is mutual
and the progenitor N$\uparrow$ is damping down P$\downarrow$ what makes it less distinctive.

Negative lags appear at frequencies higher than $\nu_{max}$ only and the difference between
the frequencies and $\nu_{max}$ is not small then. Positive lags are found at frequencies
both lower and higher than $\nu_{max}$ but mainly around $\nu_{max}$.

We normally expect that if $\gamma_i$ increases then initial $a$ decreases.
If we adopt the above assumption then, according to the received results, there is a certain 
value of the initial Lorentz factor $\gamma_i$ below which the positive lag is not visible and
with growing $\gamma_i$ positive lags become more and more harsh. If we assume additionally
that the GRB luminosity is growing together with increasing $\gamma_i$, then together with
the increase in the distance to GRBs we will be observing the larger contribution of GRBs
with stronger positive lags.

Our simulations show that the low energy electrons gain small $\beta$ if the shock decelerates
quickly (see Fig. \ref{fig:v_decel_ad} and \ref{fig:n_struct_ad}).
It is consistent with observations of the low energy electrons in plerions
(\citet{Weiler78}, $1.2<\beta<1.6$), GRBs (\citet{Band93}, $\beta\simeq1.1$ typically),
hotspots in radio galaxies (\citet{Stawarz07}, $\beta<2$) and luminous blazar sources
(\citet{Sikora09}, $\beta\simeq1.6$ typically). Moreover, the simulations show that the high
energy electrons gain $\beta$ larger than $2.23$ if the shock decelerates slowly (see Fig.
\ref{fig:v_const_ad}, $2.4<\beta<2.5$) and it is consistent with observations of the sources with
relativistic shocks where $\beta\simeq2.4$ is a typical value. Medium deceleration is presented
in Fig. \ref{fig:s_stop_ad}, there are two flat parts in spectra. Our intuitive estimation is
that the deceleration of a typical GRB is changing from $a=50$ to $a=9\cdot10^3$ during
the phenomenon.

\section{Fermi acceleration}\label{sec:Fermi}
In order to check whether Fermi acceleration is able to produce a GRB lag we have carried out
additional simulations. We have switched Cracow acceleration off through the application of $C=-2$
upstream of the shock for all simulations and we have switched Fermi acceleration on by choosing
large values of $C_d$ downstream of the shock. In this section we consider various values of $C_d$
and we use the adapted vector ($\gamma_d$, $a$, $C_d$, $sp$) instead of the old one.

At fixed values $a=10^2$ and $C_d= 5$ we carried out simulations for
$\gamma_d\in\{10\_5, 20\_10, 40\_10 \}$, and at fixed $a=10^3$ and two values of
$\gamma_d\in\{10\_5, 20\_10 \}$ we performed simulations for $C_d\in\{1, 3, 4, 5, 6, 7\}$.
It turned out that $C_d=1$ was a saturated value for the second set and increasing
it did not change the results. The results have been collected in Tables
\ref{tab:F_ad_100}-\ref{tab:F_ad_20_10}.

\begin{table}
\caption{$a=100$, spectrum - $ad$}\label{tab:F_ad_100}
\begin{center}%table
\begin{tabular}{cccrcrl}
\hline
$\gamma_d$ & $C_d$ & $n_p$ & $\nu_p$ & $n_n$ & $\nu_n$ & lag \\
\hline
$10\_5$  & 5 & 2 & 1.88 & \nodata & \nodata & P$\downarrow$ \\
$20\_10$ & 5 & 7 & 3.70 & \nodata & \nodata & P$\downarrow$ \\
$40\_10$ & 5 & 5 & 4.04 & \nodata & \nodata & P$\downarrow$ \\
\hline
\end{tabular}
\end{center}%table
\end{table}

\begin{table}
\caption{$\gamma_d$ - 10\_5, spectrum - $ad$}\label{tab:F_ad_10_5}
\begin{center}%table
\begin{tabular}{cccrcrl}
\hline
$a$ & $C_d$ & $n_p$ & $\nu_p$ & $n_n$ & $\nu_n$ & lag \\
\hline
$10^{3}$ & 1 & \nodata & \nodata & 2 & 1.97 & N$\uparrow$ \\
$10^{3}$ & 3 & \nodata & \nodata & 2 & 1.99 & N$\uparrow$ \\
$10^{3}$ & 4 & \nodata & \nodata & 2 & 2.02 & N$\uparrow$ \\
$10^{3}$ & 5 & \nodata & \nodata & 2 & 2.02 & N$\uparrow$ \\
$10^{3}$ & 6 & \nodata & \nodata & 2 & 2.02 & N$\uparrow$ \\
$10^{3}$ & 7 & \nodata & \nodata & 2 & 2.02 & N$\uparrow$ \\
\hline
\end{tabular}
\end{center}%table
\end{table}

\begin{table}
\caption{$\gamma_d$ - 20\_10, spectrum - $ad$}\label{tab:F_ad_20_10}
\begin{center}%table
\begin{tabular}{cccrcrl}
\hline
$a$ & $C_d$ & $n_p$ & $\nu_p$ & $n_n$ & $\nu_n$ & lag \\
\hline
$10^{3}$  & 1 & 1 & 3.68 & \nodata & \nodata & P$\downarrow$ \\
$10^{3}$  & 3 & 1 & 3.62 & \nodata & \nodata & P$\downarrow$ \\
$10^{3}$  & 4 & 1 & 3.63 & \nodata & \nodata & P$\downarrow$ \\
$10^{3}$  & 7 & 1 & 3.66 & \nodata & \nodata & P$\downarrow$ \\
\hline
\end{tabular}
\end{center}%table
\end{table}

\begin{figure}
\centering{\includegraphics[width=78mm]{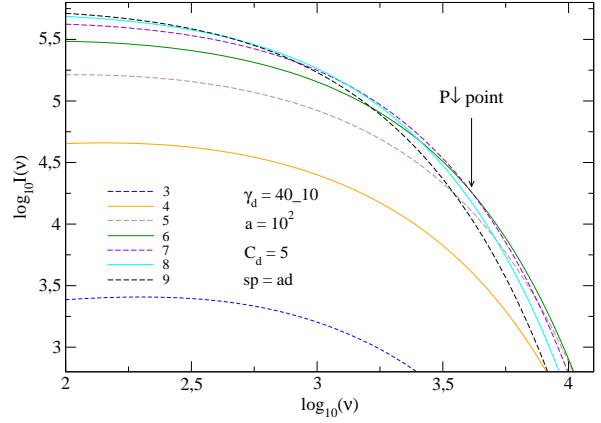}}
%\begin{center}%figure_7_1
%\FigureFile(78mm,){ref_lag_ad.eps}
%\end{center}%figure_28
\caption{Fermi acceleration. Photon spectra of (40\_10, $10^2$, 5, ad).
         Fermi acceleration does not influence P$\downarrow$ produced by reflection process.}
\label{fig:ref_lag_ad}
\end{figure}

\begin{figure}
\centering{\includegraphics[width=78mm]{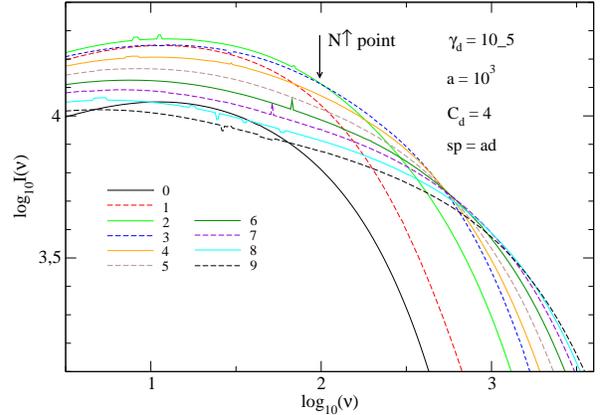}}
%\begin{center}%figure_7_2
%\FigureFile(78mm,){fermi_lag_ad.eps}
%\end{center}%figure_29
\caption{Fermi acceleration. Photon spectra of (10\_5, $10^3$, 4, ad). Fermi acceleration
         generates N$\uparrow$ and reflection process is too poor to suppress this.}
\label{fig:fermi_lag_ad}
\end{figure}

We have checked spectra 'rd' and 'pd' and it turns out that P$\downarrow$ is a pure outcome of
reflection process because spectra 'pd' do not show a lag in this case
(Fig. \ref{fig:ref_lag_ad}). The negative lag arises in full measure from Fermi
acceleration what is visible in spectra 'pd'. In this case spectra 'rd' are showing
P$\downarrow$ which is suppressed by N$\uparrow$ from spectra 'pd' and it gives spectra
'ad' with N$\uparrow$ (see Fig. \ref{fig:fermi_lag_ad}).

Findings are simple. Fermi acceleration generates negative lags only in shocks that
have small Lorentz factors and are slowing down slowly. In other conditions positive lags
produced by reflection process are visible.

\begin{figure}
\centering{\includegraphics[width=78mm]{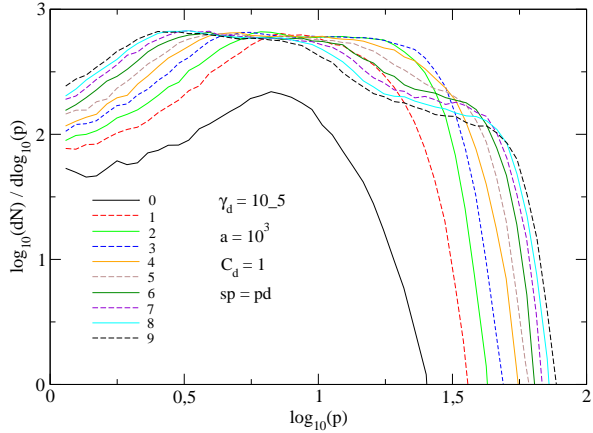}}
%\begin{center}%figure_7_3
%\FigureFile(78mm,){fermi_acc_pd.eps}
%\end{center}%figure_30
\caption{Fermi acceleration. Particle spectra of (10\_5, $10^3$, 1, pd). The spectra are similar
         to the corresponding spectra with active Cracow acceleration producing N$\uparrow$.}
\label{fig:fermi_acc_pd}
\end{figure}

\begin{figure}
\centering{\includegraphics[width=78mm]{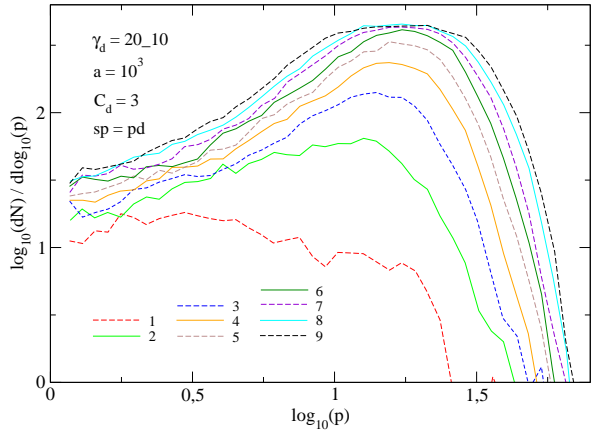}}
%\begin{center}%figure_7_4
%\FigureFile(78mm,){fermi_off_pd.eps}
%\end{center}%figure_31
\caption{Fermi acceleration. Particle spectra of (20\_10, $10^3$, 3, pd).
         The corresponding photon spectra do not show neither negative nor positive lags.}
\label{fig:fermi_off_pd}
\end{figure}

The first question is which way the negative lag is produced. We have not prepared an expert
opinion on this issue, but it has emerged a simple interpretation after analysing particle
spectra 'pd' (Fig. \ref{fig:fermi_acc_pd}, \ref{fig:fermi_off_pd}). Spectra in Fig.
\ref{fig:fermi_acc_pd} are similar to the corresponding spectra with active Cracow acceleration.
The spectra are falling at medium particle energies and moving forward at high-energy tail.
In the process of Fermi acceleration particles are wandering UP over a distance shorter than
in Cracow acceleration, but for small Lorentz factors of the shock it is enough to produce
lingering of the high-energy tail in the particle spectra.


The second question is why there is no negative lag for ($10\_5$, $100$, $5$, ad).
It results probably from the fact that Fermi acceleration has longer acceleration time and produces
steeper spectra than Cracow acceleration. It explains the lack of negative lags
in photon spectra of rapidly slowing down shocks.

The third question is why one can not see negative lags for larger Lorentz factors of the shock.
The answer is that the wandering time UP decreases more quickly with the growing Lorentz factor
of the shock than the time in Cracow acceleration or Fermi acceleration fades out.

On the basis of the received results we put forward the hypothesis that the negative lag
can be produced by any acceleration mechanism in which particles after reflection process
are crossing the shock repeatedly from UP to DOWN and around DOWN to UP. 

We have not examined Fermi acceleration in detail and we think that the Weibel instability and
the fact that particles are accelerating in PWNe make this mechanism impossible.
Our numerical code is still relatively straightforward and therefore it can not be ruled out on
the basis of the current results that the negative lag arises due to Fermi acceleration. More
detailed simulations and observations are needed in order to decide conclusively on the matter.

\section{Two acceleration processes}\label{sec:two_processes}
Observations of GRBs with \textit{the Fermi Gamma Ray Space Telescope} indicate that photons
in the GeV range (the LAT) and photons in the MeV range (the GBM) arrive at different times.
The GeV photons arrive sometimes later and sometimes before than the MeV photons
(\citet{Abdo09}, \citet{Ghirlanda10}). It seems that the observed spectra in the two ranges
originate from different regions. However, it is the very same phenomenon that occurs in one
region and we will describe it below.

In order to understand this phenomenon one should realise that Cracow acceleration does not have
a general restriction to the maximum energy achieved by particles. The restrictions result from
specific physical conditions only. Values of $C$ and $\gamma$ are two basic parameters that are
limiting the energy of accelerated particles. Increasing any of them increases this limit.

The acceleration time is another important quantity and although it is constant downstream
of the shock, it is $\gamma$ times smaller in the upstream plasma rest frame and grows together
with reducing $\gamma$.

\textit{The Fermi Telescope} observes bursts which have sufficiently large values of $C$ and
$\gamma$ to allow for Cracow acceleration to take place at GeV energies detected on Earth. When
\textit{the Fermi Telescope} is detecting the onset of the GeV emission then the accelerated
particles achieve energies enabling them to produce synchrotron emission in this range. 

The two acceleration processes behave differently during the phenomenon.
Reflection process is weakening almost all the time except for the short initial time when
the number of particles entering the process is larger than the number of particles coming out.
The weakening manifests itself by the moving of the maximum of the energy distribution of
accelerated particles toward lower energies. We will call the maximum the reflection surge.

Cracow acceleration accelerates particles all the time but unevenly. The acceleration time
decreases and this influences each particle the same. Reflection process provides particles
with energy that is more and more lower, therefore Cracow acceleration accelerates more and more
smaller number of particles at a fixed particle energy. That does not concern particles with
the highest energies.

We expect that the real equation of motion of the shock is much more dynamic than that in our
simulations. The value of $a$ increases with decreasing $\gamma$. The shock with a small value
of $a$ has more particles UP than the shock with a large value of $a$. During the rapid phase
the shock produces a large amount of high-energy particles UP. The particles make up
an additional fraction of non-thermal particles UP. It is convenient to regard the fraction
as seed particles produced in the rapid phase. We will call the fraction the rise-fraction.

When the shock slows down and $a$ increases then the Cracow acceleration takes up
the rise-fraction. This produces a weakening wave moving toward higher energies in the energy
distribution of accelerated particles. We will call the wave the Cracow surge.

The question is why we can not see the Cracow surge in the present simulations. It results
from the simplicity of the equation of motion which does not allow for a large dynamics.
The simulations show the lags because they are strong local phenomena. The Cracow surge is
a global phenomenon, but there are some signs of it in the present calculations.

One can see bumps at high energies in 'ad' particle spectra where the Cracow surge should occur.
At later times the bumps move toward low energies, but it is also caused by the low dynamic range.
The corresponding 'ad' photon spectra show an extended plateau instead of the maximum, but better
figures have been not presented in the paper. In real conditions the rise-fraction is produced
quickly (small $a$) and then Cracow acceleration processes it slowly (large $a$).

A GRB starts at the energy of accelerated particles that is equal to the energy of reflected
particles at the onset of the phenomenon. A short time after the onset two bulges appear in
the energy spectrum, the reflection surge moves toward low energies and the Cracow surge moves
toward high energies. The transition of a surge is seen as a signal. Each of two detectors
of photons working in different spectral bands can detect one signal only. The sequence
of the detections depends on the specific GRB.

There is the option that an instrument could detect the reflection surge at the very onset
of a GRB. Shortly after the detection the same instrument detects the Cracow surge. The light
curve produced by the instrument will exhibit the structure with two pulses (the double pulse).
Such structures are present in \textit{BATSE} data.

We describe the Cracow surge in more detail below. We will use the observation of GRB 090510
(\citet{Ghirlanda10}) as the example.

GRB 090510 was observed at MeV energies at first. This pulse could originate both from
the reflection surge as well as from the Cracow surge. The pulse at GeV energies originates
from the Cracow surge. The period of time between the pulses is considerable, therefore we can
assume that during the GeV pulse the value of $a$ is large and constant. The rise-fraction
is present upstream of the shock during the pulse.

The rise-fraction consists of particles that are more close to the shock and have lower energies
and larger anisotropy, and farther with higher energies and smaller anisotropy. The anisotropy 
does not play a significant role in the case. The acceleration time changes very slowly for
large $a$ and we can treat it as constant.

At first, the shock is getting the particles of the rise-fraction with lower energies and
accelerates them. When the shock reaches the particles with higher energies, then the lower-energy
particles were in time to accelerate to energies comparable to the higher-energy particles.
As a result instead of the value of $\beta$ typical for given $a$, e.g. $\beta=2.6$, a flatter
particle spectrum is arising, e.g. $\beta=1.4$, which will approach $\beta=2.6$ going to higher
energies, but will enter the GeV photon spectrum range with $\beta=1.6$ for example. Next,
the normal seed particles, that produce $\beta=2.6$, are falling within the GeV range and
the photon spectrum starts falling.

The above description is compatible with the observation of GRB 090510. At the high-energy
part of the photon spectrum the photon index is equal to $\beta/2+1$. In the GeV range
the photon spectrum at first grows with the photon index equal to $1.8$ ($\beta=1.6$), and then
it is falling with the photon index equal to $2.3$ ($\beta=2.6$). 

\section{UHECRs acceleration}\label{sec:UHECRs}

We claim that the mechanism of Cracow acceleration produces UHECRs.
The restrictive requirement that rules out UHECRs acceleration in most astrophysical
objects is that the particle gyroradius must be much smaller than the system size. The restriction
does not exist if the acceleration takes place at standing or moving with a mildly relativistic
velocity shocks (with respect to ISM), UP has the Lorentz factor larger than $\sim 1000$
(with respect to ISM) and UHECRs come out from UP. Two kinds of sources fulfil the requirements,
PWNe and relativistic flows in AGN jets. The PWN sources overcome the restriction of
the Greisen-Zatsepin-Kuzmin cutoff also.

Below, we explain why Cracow acceleration is able to accelerate UHECRs. All quantities are given
in the plasma rest frame. Let $E_u$ be the particle energy UP, $E_d$ the particle energy DOWN,
$B_u$ the homogeneous magnetic field UP and $B_d$ the homogeneous magnetic field DOWN. When
a particle crosses the shock from UP to DOWN then its energy changes according to the formula
$E_d=E_u/\gamma_{ud}$. When a particle crosses the shock from DOWN to UP then its energy changes
according to the formula $E_u=2\gamma_{ud}E_d$. This is a result of the particle anisotropy
upstream of the shock. The relation between magnetic fields is $B_d=4\gamma_{ud}B_u$. We assume
that DOWN is at rest or moves with the velocity $\lesssim 0.95c$ with respect to ISM.

Synchrotron losses are $\sim E^2B^2$. It gives that the synchrotron losses of the particle that
crossed the shock from DOWN to UP are $\sim 4$ times slower than the losses before the crossing.
The particle spends much shorter time UP than DOWN so that the synchrotron losses UP can be
neglected. Synchrotron losses of protons and nuclei are neglected DOWN and their achieved maximum
energies DOWN depend on the size of the system only. 

The accelerated particles do not escape from DOWN if the perpendicular size of system is larger
or equal to the particle gyroradius $r_g$. It yields the formula for the maximum energy DOWN,
$E_d\simeq 300d_dB_d\; [eV]$, where $d_d\; [cm]$ is the size of DOWN perpendicular to the shock
normal, $B_d\; [Gs]$. Protons and nuclei are able to reach energies equal to
$\sim 10^{14} - 10^{16} eV$ downstream of the shock and it is enough to produce UHECRs if
the Lorentz factor of the shock $\gamma\gtrsim 10^3$ and the particles get out into ISM through UP.

We assume that a particle crossed the shock from DOWN to UP and its momentum was parallel
to the shock normal at the beginning of the wandering UP. After a while the direction
of the particle momentum changes and $\alpha$ is the angle between the momentum and
the shock normal. The angle $\alpha$ is small therefore the distance the particle
covered in the direction perpendicular to the shock normal is $d_u\simeq r_g\alpha^2/2$,
where $r_g$ is the particle gyroradius UP. The value of $d_u$ is the same both
in the upstream and in the downstream plasma rest frame but the particle gyroradius
grows $\sim 8\gamma_{ud}^2$ times when the particle crosses the shock from DOWN to UP.
The maximum perpendicular (to the shock normal) length of the particle path DOWN is
$d_d\simeq r_g$ and therefore $d_u/d_d\simeq 4\gamma_{ud}^2\alpha^2$.

We will estimate the value of $\alpha$ in two ways. First, for $\beta$ slightly larger
than $\beta_0$, we apply $\alpha\simeq\Delta t_u$. We eliminate $Q_0$ from
(\ref{eq:with_g}), (\ref{eq:dtu}) and get $\Delta t_u\simeq 0.51(\beta-2.23)^{0.695}/\gamma$.
If we put the values $\beta=2.3, 2.4, 2.6$ into the equation we get
$d_u/d_d\simeq 0.013, 0.045, 0.13$.

The electromagnetic forces that disturb the particle movement in the homogeneous magnetic field
twist its path with different radii of curvature. If we exclude the contribution of the homogeneous
magnetic field then we get the change of the direction of momentum and displacement of the particle
caused by a scattering and, in consequence, the scattering radius. We will use the parameter $\eta$
that is the ratio of the particle gyroradius to the mean scattering radius.

In our simulations $\eta$ is equal to infinity. In real plasma $\eta$ has a finite value. If $\eta$
takes a finite value then the first estimation of $d_u/d_d$ becomes larger for half of
the particles. Because $d_u/d_d\sim\alpha^2$, the first estimation gives that the deceleration of
the shock must increase $\alpha$ a few times to allow accelerated particles come out from UP.

The second estimation is obtained in the case of the neglected influence of the homogeneous
magnetic field. The angle $\alpha$ results from scatterings only and $\beta=\beta_0$. We apply
$\alpha\simeq\Delta\alpha_u$, where $\Delta\alpha_u\simeq1.3/\gamma$
(see Fig. \ref{fig:time_angle}), but we divide the result by $\eta$ because the particle turns in
electromagnetic disturbances. It yields the second estimation $d_u/d_d\simeq 3.4/\eta$. We suggest
that loop plasma has the value of $\eta$ in the range $10-100$. It gives that if $\alpha$ increases
$\sim2$ times up to a few times because of the deceleration of the shock then the particles
come out from UP.

In the model of loop plasma the homogeneous magnetic field and electromagnetic curvatures are
generated by electric current loops. The loops could change if $\gamma$ changes. We think that
$B_u$ could decrease with decreasing $\gamma$. It would explain, at least partly, dissipation
of the Poynting flux of the pulsar wind before it reaches the inner edge of the Crab Nebula.

A few times increase in the value of $\alpha$ as the result of the shock deceleration is something
expected. The slowing down velocity of the shock enables particles to spend longer time UP and it
can not be neglected. If $B_u$ decreases then $r_g$ increases and, as a result, $d_u$ increases and
the particle spends longer time UP. It could even produce the feedback that prevents the particle
from returning to the shock. However, the reduction of the shock velocity can not be too large
because the energy $E_d$ increases $\sim2\gamma_{ud}\gamma_{is}$ times, where $\gamma_{is}$ is
the Lorentz factor of the plasma UP as seen in ISM at the moment when the particle
comes out from UP.

Loop plasma has local electric fields that are perpendicular to the shock normal in a statistical
way. It means that disturbances of the particle movement UP could act to increase the value of
$d_u/d_d$ that was estimated earlier. We do not know how the fields influence the value, but
the fields could be so strong that $d_u/d_d$ is larger than 1 for constant velocity shocks with
a sufficient value of $\gamma$.

The presented calculations are promising but further numerical simulations are needed to show that
Cracow acceleration accelerates UHECRs. The first kind of simulations will be similar to presented
in this paper but the numerical code will include an extended equation of motion and variable
$B_u$. The second kind of simulations should search for values of $\eta$. One must choose
parameters of loop plasma and follows particle trajectories in its electromagnetic field. The third
kind of simulations will be similar to the first kind but UP will be replaced with loop plasma.

If Cracow acceleration accelerates UHECRs then astronomical observations will show that UHECRs
arrive from systems where upstream plasma moves away from an observer and is slowing down from
$\gamma\gtrsim10^3-10^2$ to $\gamma\gtrsim10^2$.

Fermi acceleration can not accelerate UHECRs because its change of the direction
of the particle momentum that enables the return to the shock takes place DOWN.
For example, \citet{Reville14} disfavour weakly magnetised ultrarelativistic shocks,
in context of Fermi acceleration, as high-energy particle accelerators.

Rounding off this section, we address the problem of UHECRs events composition.
The particle gyroradius $r_g\sim E/e$, where $E$ is the particle energy and $e$ is its
charge. It implies that at the same energy heavy nuclei (Fe, Ni) have the smallest $r_g$.
On that account, we divide slowing down ultrarelativistic shocks into two categories.

The first kind are shocks that slow down so slowly that the heavy nuclei are able
to escape from UP. They produce the spectrum where the atomic number decreases with
decreasing energy.

The second kind are shocks that slow down rapidly. They can be divided into two
subclasses. The shocks of the first subclass slow down according to a function
that is more linear and the second subclass according to a function that is more
exponential. The first subclass produces the spectrum where the atomic number
decreases with decreasing energy, but the highest energy particles are lighter
nuclei (O, N for example). The second subclass, because of much smaller energy gain
during second phase of the more exponential deceleration, forms two components
and the atomic number in each component decreases with decreasing energy.
For example, the elements make a sequence Be, He, H, O, N, B if their energies
decrease.

In general, it should be noted that the first kind of shocks are able to produce
particles that have higher energies than the second kind. The rule is the same
in the subclasses of the second kind. It implies that Cracow acceleration prefers
heavy or light nuclei to be UHECRs and not protons. The problem, from
the observational point of view, is presented in \citet{Taylor14} and 
\citet{Fargion15}.

\section{Seed particles}\label{sec:seed_particles}
In our model presented in \citet{Bednarz04} we proposed some acceleration mechanisms to
be responsible for production of seed particles and that their radiation is detected as
the precursor of GRBs. Actually, the mechanisms are ineffective in pre-acceleration of protons
and nuclei and the analysis of spectra of GRB precursors is suggesting that they are produced
in the same mechanism as GRBs. In our opinion, the mechanism described by \citet{Hoshino92} is
the only (among the proposed) effective mechanism that is able to produce seed leptons.
This is the process of the downstream plasma acceleration of positrons to non-thermal
distributions. Cracow acceleration needs seed particles upstream of the shock to accelerate them.
We propose two ways in which they could be injected into UP.

The first way occurs in the relativistic jets that are fired from a massive object close to
the neighbourhood. The jet production mechanism and indeed the jet composition on very small
scales are not known at present. We propose that seed particles are present in jets from
the beginning of producing them and are there all the time. We think that loop plasma generates
seed particles both leptons and nuclei in its local electric fields. The fields are
perpendicular to the shock normal in a statistical way and could produce an anisotropic
distribution of seed particles.

The second way is at work in a pulsar wind when it is streaming into the ambient medium and
creating a standing shock. Seed particles can, similarly as in jets, be present from the moment
of generating the wind. However, it concerns leptons only. We propose that seed particles,
particularly protons and nuclei, are entering the pulsar wind from its side boundary. These
particles are coming out from the ambient medium being pushed by the gravitational force of
the pulsar. Their Lorentz factor in the wind plasma rest frame is equal to $\sim\gamma_w$,
where $\gamma_w$ is the Lorentz factor of the pulsar wind, so that they become seed
particles automatically.

\section{Summary and Discussion}\label{sec:summary}
Up to now, simulations of particle acceleration at relativistic shocks concerned constant
velocity shocks only. We are the first who have simulated particle acceleration at slowing down
relativistic shocks and that is why we have solved the mysteries of photon and particle spectra
of the astronomical objects that radiate the synchrotron radiation of particles accelerated
at relativistic shocks. No one was trying to perform such simulations before because it was a very
difficult task. It turns out that GRB lags and the hard low energy spectral indices arise due to
the deceleration (section \ref{sec:results}).

We have presented the mechanism of Cracow acceleration (section \ref{sec:Cracow}). The mechanism
increases its strength when the Lorentz factor of the shock grows and initial $\beta$
does not depend on the Lorentz factor but on the conditions of the upstream plasma.

We have shown that the whole acceleration consists of two processes, reflection process and Cracow
acceleration. We have assumed that the real deceleration is more dynamic than that in our
simulations and the rise-fraction of particles upstream of the shock if formed at the strong
deceleration stage. Acceptance of the assumption gives explanation to the question why
two transitions are observed at one GRB and why the observation at high energy can follow
the observation at low energy and vice versa (section \ref{sec:two_processes}). It also explains
why the photon spectrum is hard at the rise stage of the Cracow surge. Sometimes, two surges
produce the double pulse, the first maximum in the pulse originates from the reflection surge and
the second from the Cracow surge.

Cracow acceleration requires strong disturbances of the movement of particles upstream of
the shock in order to accelerate the particles. We have shown that current PIC simulations of
relativistic plasma are useless. We have presented the model of the microphysics of relativistic
flows that would allow for acceleration in the mechanism of Cracow acceleration
(section \ref{sec:micro}). We have sketched the PIC simulations that are needed to solve
the problem of the microphysics of relativistic plasma. The model of loop plasma gives
the homogeneous magnetic field that is perpendicular to the shock normal but an external magnetic
field, for example the magnetic field of a central object, could change this geometry.

The current PIC simulations suggest that seed particles could be produced in the upstream plasma.
We agree with the idea, but only proper PIC simulations can prove this. We proposed that 
particles of the ambient medium are injected into pulsar wind of PWNe and become seed particles
(section \ref{sec:seed_particles}).

We claim that Cracow acceleration produces UHECRs which come out of UP
(section \ref{sec:UHECRs}). Relativistic shocks are present in supernovae (SNe) and PWNe
which are supposed to be the sources of cosmic rays in the Galaxy. They are also present
in relativistic flows ejected from ANGs. The astronomical phenomena are widespread and show
synchrotron radiation produced by energetic particles. The accelerated particles have power-law
spectra similar to cosmic rays. The energy of these shocks is the main energy of the system.
There are no other reasonable sources of cosmic rays but relativistic shocks.
We claim that reflection process and Cracow acceleration produce almost all cosmic rays
with energies higher than $\sim 50$ GeV.

This paper is almost the same as \citet{Bednarz12}. At the beginning of 2016 four paragraphs
were added at the end of section \ref{sec:UHECRs}, the bibliography has been updated and
the paper has been sent to {\itshape Reports on Progress in Physics}. This paper is the same
as that version. The journal did not consider it because it did not meet the requirements
of a review article.

\end{document}